\documentclass[11pt]{article}
\usepackage{amsmath,amssymb}
\usepackage[usenames]{color}
\usepackage{pstricks}
\numberwithin{equation}{section}

\newcommand{\nc}{\newcommand}
\def\vvdots{\mathinner{\mkern1mu\raise1pt\vbox{\kern7pt\hbox{.}}\mkern2mu
  \raise4pt\hbox{.}\mkern2mu\raise7pt\hbox{.}\mkern1mu}}
\nc{\fh}{\hat{f}}
\nc{\muh}{\hat{\mu}}
\nc{\nuh}{\hat{\nu}}
\nc{\bib}{\bibitem}
\nc{\al}{\alpha}
\nc{\g}{\gamma}
\nc{\G}{\Gamma}
\nc{\D}{\Delta}
\nc{\eps}{\epsilon}
\nc{\la}{\lambda}
\nc{\La}{\Lambda}
\nc{\var}{\varphi}
\nc{\pa}{\partial}
\nc{\nn}{\nonumber \\ }
\nc{\hf}{\frac{1}{2}}
\nc{\dz}{\frac{dz}{2\pi i}}
\nc{\bin}[2]{\left(\!\!\!\begin{array}{c} {#1}\\ {#2} \end{array}\!\!\!\right)}
\nc{\be}{\begin{equation}}
\nc{\ee}{\end{equation}}
\nc{\bea}{\begin{eqnarray}}
\nc{\eea}{\end{eqnarray}}
\nc{\bra}[1]{\langle {#1}|}
\nc{\ket}[1]{|{#1}\rangle}
\nc{\ketw}[1]{({#1})^{\phantom{a}}_{{\cal W}}}
\nc{\chit}{\raisebox{0.25ex}{$\chi$}}
\nc{\chih}{\raisebox{0.25ex}{$\hat\chi$}}
\nc{\Db}{\mbox{\boldmath $D$}}
\nc{\Hb}{\mbox{\boldmath $H$}}
\nc{\Hc}{\mathcal{H}}
\nc{\Lc}{\mathcal{L}}
\nc{\Nc}{\mathcal{N}}
\nc{\Rc}{\mathcal{R}}
\nc{\Vc}{\mathcal{V}}
\nc{\Ib}{\mbox{\boldmath $I$}}
\nc{\qb}{\bar{q}}
\nc{\Jh}{\hat{J}}
\nc{\Qh}{\hat{Q}}
\nc{\vh}{\hat{v}}
\nc{\wh}{\hat{w}}
\nc{\Jt}{\tilde{J}}
\nc{\Qt}{\tilde{Q}}
\nc{\vt}{\tilde{v}}
\nc{\wt}{\tilde{w}}
\nc{\vb}{\mathbf{v}}
\nc{\Ac}{\mathcal{A}}
\nc{\Bc}{\mathcal{B}}
\nc{\Cc}{\mathcal{C}}
\nc{\Dc}{\mathcal{D}}
\nc{\Ec}{\mathcal{E}}
\nc{\Fc}{\mathcal{F}}
\nc{\Ic}{\mathcal{I}}
\nc{\Jc}{\mathcal{J}}
\nc{\Oc}{\mathcal{O}}
\nc{\Qc}{\mathcal{Q}}
\nc{\Sc}{\mathcal{S}}
\nc{\Xc}{\mathcal{X}}
\nc{\Yc}{\mathcal{Y}}
\nc{\Zc}{\mathcal{Z}}
\nc{\fus}{\mbox{}\,\hat\otimes\,\mbox{}}
\nc{\Sct}{\tilde{\mathcal{S}}}
\nc{\ch}{{\rm ch}}
\nc{\R}{{\cal R}}
\nc{\dkk}{\delta_{j,\{k,k'\}}^{(2)}}
\nc{\ddkk}{\delta_{j,\{k,k'\}}^{(4)}}
\nc{\dddkk}{\delta_{j,\{k,k'\}}^{(8)}}
\nc{\dnn}{\delta_{j,\{n,n'\}}^{(2)}}
\nc{\ddnn}{\delta_{j,\{n,n'\}}^{(4)}}
\nc{\dddnn}{\delta_{j,\{n,n'\}}^{(8)}}
\def\vvdots{\mathinner{\mkern1mu\raise1pt\vbox{\kern7pt\hbox{.}}\mkern2mu
  \raise4pt\hbox{.}\mkern2mu\raise7pt\hbox{.}\mkern1mu}}
\nc{\gauss}[2]{\left[\!\!\begin{array}{c} {#1}\\ {#2} \end{array}\!\!\right]}
\nc{\sbin}[2]{\left\{\!\!\!\begin{array}{c} {#1}\\ {#2} 
\end{array}\!\!\!\right\}}
\nc{\sbinlr}[2]{\Big\langle\!\!\begin{array}{c} {#1}\\ {#2} 
\end{array}\!\!\Big\rangle}
\nc{\bino}[2]{\left(\!\!\begin{array}{c} {#1}\\ {#2} \end{array}\!\!\right)}
\def\half {\mbox{$\textstyle \frac{1}{2}$}}

\definecolor{lightblue}{rgb}{.61,.61,1}
\definecolor{midblue}{rgb}{.7,.7,1}
\definecolor{lightlightblue}{rgb}{.85,.85,1}
\definecolor{lightestblue}{rgb}{.96,.96,1}

\begin{document}

\topmargin -5mm
\oddsidemargin 5mm

\setcounter{page}{1}

\mbox{}\vspace{-16mm}
\thispagestyle{empty}

\begin{center}
{\LARGE {\bf Fusion matrices, generalized Verlinde formulas,}}\\[.2cm]
{\LARGE {\bf and partition functions in ${\cal WLM}(1,p)$}}

\vspace{7mm}
{\Large J{\o}rgen Rasmussen}
\\[.3cm]
{\em Department of Mathematics and Statistics, University of Melbourne}\\
{\em Parkville, Victoria 3010, Australia}\\[.4cm]
{\tt J.Rasmussen\!\!\ @\!\!\ ms.unimelb.edu.au}

\end{center}

\vspace{8mm}
\centerline{{\bf{Abstract}}}
\vskip.4cm
\noindent
The infinite series of logarithmic minimal models ${\cal LM}(1,p)$ is considered in 
the ${\cal W}$-extended picture where they are denoted by ${\cal WLM}(1,p)$. 
As in the rational models, the fusion algebra of
${\cal WLM}(1,p)$ is described by a simple graph fusion algebra. 
The corresponding fusion matrices are mutually commuting, but in general not diagonalizable. 
Nevertheless, they can be simultaneously brought to Jordan form 
by a similarity transformation.
The spectral decomposition of the fusion matrices is completed by
a set of refined similarity matrices converting the fusion matrices into Jordan
{\em canonical} form consisting of Jordan blocks of rank 1, 2 or 3.
The various similarity transformations and Jordan forms are determined
from the modular data. This gives rise to a generalized Verlinde formula for the fusion matrices.
Its relation to the partition functions in the model is discussed in a general framework.
By application of a particular structure matrix and its Moore-Penrose inverse,
this Verlinde formula reduces to the generalized Verlinde formula for 
the associated Grothendieck ring.
\vskip1cm
%
%
\renewcommand{\thefootnote}{\arabic{footnote}}
\setcounter{footnote}{0}

\section{Introduction}
\label{SecIntro}

The fusion matrices of a standard rational conformal field theory are diagonalizable.
This is made manifest by the Verlinde formula~\cite{Ver88,DiFMS} where the diagonalizing
similarity matrix is the modular $S$-matrix of the characters in the theory.
In a logarithmic conformal field theory, on the other hand, there are typically 
more linearly independent representations
than linearly independent characters due to the presence of indecomposable representations
of rank greater than 1. Consequently, there is no Verlinde formula in the usual sense
and the fusion matrices may not all be diagonalizable. This is indeed the situation
in the cases studied here, but can be circumnavigated in some logarithmic models
where a Verlinde formula is recovered when restricting to a subset of the spectrum of 
representations and their associated characters~\cite{PR0902}.
Our results here present the first spectral decompositions of non-diagonalizable fusion matrices.
Spectral decompositions of the likewise non-diagonalizable matrix realizations of the 
associated Grothendieck rings appear in~\cite{PRR0907}, see below. 

We consider the infinite series of logarithmic minimal models ${\cal LM}(1,p)$~\cite{PRZ0607}
in the ${\cal W}$-extended picture~\cite{PRR0803} where they are denoted by ${\cal WLM}(1,p)$.
The fusion rules~\cite{GK9606,EF0604,GR0707,PRR0803} underlying the commutative
and associative fusion algebra of ${\cal WLM}(1,p)$ are generated from repeated fusions
of the two fundamental representations $\ketw{2,1}$ and $\ketw{1,2}$.
As in rational conformal field theories~\cite{DiFMS}, the fusion algebra is described by a 
simple graph fusion algebra. This is neatly encoded in the graphs associated to the 
two fundamental fusion matrices, and we exhibit these graphs explicitly.

There are $4p-2$ indecomposable representations in the model ${\cal WLM}(1,p)$.
According to~\cite{Ras0812}, every associated fusion matrix 
can be written as a polynomial in the fundamental fusion matrices
$\Nc_{\ketw{2,1}}$ and $\Nc_{\ketw{1,2}}$.
We devise a similarity transformation in the form of a matrix $Q$ which 
converts these two fusion matrices simultaneously into Jordan canonical form. 
This matrix $Q$ is naturally described in terms of Chebyshev
polynomials and derivatives thereof. It is constructed by concatenating a complete set
of generalized eigenvectors of $\Nc_{\ketw{1,2}}$ forming Jordan chains of length 1 or 3.

Due to the polynomial constructions just mentioned, the remaining fusion matrices are
also brought to Jordan form by the similarity matrix $Q$, albeit 
typically {\em non-canonical} Jordan forms. 
The similarity matrices converting them into {\em canonical} Jordan forms can be obtained 
rather straightforwardly from $Q$.
For every fusion matrix $\Nc$, we thus provide this modified $Q$-matrix $Q_\Nc$ as
well as the corresponding Jordan canonical form of $\Nc$. 
Only Jordan blocks of rank 1, 2 or 3 appear in these Jordan canonical forms.

One can associate a logarithmic generalization of the Verlinde formula to 
the so-called Grothendieck ring of ${\cal WLM}(1,p)$~\cite{GR0707,PRR0907}.
This formula yields matrix realizations of the Grothendieck generators, of which
there are $2p$, in terms of the (generalized) $S$-matrix. The Jordan forms of these 
matrices contain non-trivial Jordan blocks of rank 2, while the {\em fusion} 
matrices of ${\cal WLM}(1,p)$ also 
give rise to Jordan blocks of rank 3, as already mentioned. We also stress that there are almost
twice as many representations ($4p-2$) than Grothendieck generators ($2p$), the latter number
being equal to the number of linearly independent characters.
The coincidence of the Grothendieck ring with the fusion algebra in a standard conformal
field theory therefore fails to extend to the logarithmic conformal field theory ${\cal WLM}(1,p)$.

Another objective of the present work is to generalize the results of~\cite{GR0707,PRR0907} 
by expressing the fusion matrices in terms of the modular data encoded in the generalized
$S$-matrix. This yields a Verlinde-type formula for the fusion matrices themselves and not 
just for the generators of the associated Grothendieck ring. 
Other approaches to a Verlinde formula for the ${\cal WLM}(1,p)$ models have been
proposed in~\cite{FHST0306,FK0705,GT0711}. 
As discussed in~\cite{PRR0907}, however, they do not seem to
be equivalent to the approach used in~\cite{PRR0907} which is adopted here.

Finally, we outline a general framework within which it makes sense to discuss rings 
of equivalence classes of fusion-algebra generators. Specializing this to the equivalence classes 
obtained by elevating the character identities of ${\cal WLM}(1,p)$ to equivalence 
relations between the corresponding fusion generators, we recover
the Grothendieck ring of ${\cal WLM}(1,p)$. From the lattice description of ${\cal WLM}(1,p)$,
the $4p-2$ indecomposable representations mentioned above are naturally associated with
boundary conditions. As already indicated, the corresponding characters are not linearly
independent, but we can nevertheless talk about partition functions arising
when combining two such boundary conditions. This provides a direct relationship
between the generalized Verlinde formulas for the fusion algebra and the Grothendieck
ring, respectively. The structure matrix, which governs the expansion of the reducible
characters (of the rank-2 representations) in terms of the irreducible characters, can subsequently
be used to express the Grothendieck (Verlinde) matrices in terms of the fusion (Verlinde)
matrices. This explicit relation also involves the Moore-Penrose inverse of the structure matrix.

A logarithmic minimal model ${\cal LM}(p,p')$ is defined for every pair of relatively
prime integers $1\leq p<p'$~\cite{PRZ0607}.
Its ${\cal W}$-extended picture ${\cal WLM}(p,p')$ is described 
in~\cite{PRR0803,RP0804,Ras0805},
including the fusion algebra of the set of indecomposable representations naturally
associated with boundary conditions. 
In the models ${\cal WLM}(p,p')$ with strict inequality $1<p$, 
however, there are additional irreducible representations whose
fusion properties have been systematically examined only very 
recently~\cite{GRW0905,Ras0906,Wood0907}.
It would be of interest to extend the work presented here and in~\cite{GR0707,PRR0907}
on the series ${\cal WLM}(1,p)$ to the general series ${\cal WLM}(p,p')$.

\section{Logarithmic minimal model ${\cal WLM}(1,p)$}
\label{SecWLM}

There is a ${\cal W}$-extended logarithmic minimal model ${\cal WLM}(p,p')$ for
every co-prime pair of positive integers $p<p'$~\cite{Ras0805}. Since our interest here is in 
the series of these models with first label equal to 1, we write ${\cal WLM}(1,p)$, for simplicity,
where $p>1$.

The model ${\cal WLM}(1,p)$ consists of
$2p$ ${\cal W}$-irreducible representations $\ketw{\kappa,s}$
and $2p-2$ ${\cal W}$-indecomposable rank-2 representations 
$\ketw{\R_{\kappa}^{b}}=\ketw{\R_{\kappa,p}^{0,b}}$. The set of these
${\cal W}$-indecomposable representations is
\be
 \Ic\ =\ \big\{\ketw{\kappa,s},\ketw{\R_{\kappa}^{b}};\  \kappa\in\mathbb{Z}_{1,2},
     s\in\mathbb{Z}_{1,p}, b\in\mathbb{Z}_{1,p-1}\big\}
\label{Jp}
\ee
and has cardinality $4p-2$. Here we have introduced
\be
 \mathbb{Z}_{n,m}\ =\ \mathbb{Z}\cap[n,m],\qquad\quad n,m\in\mathbb{Z}
\ee 
and unless otherwise specified, we let 
\be
 \kappa,\kappa'\in\mathbb{Z}_{1,2},\qquad\quad
   s,s'\in\mathbb{Z}_{1,p}, \qquad\quad b,b'\in\mathbb{Z}_{1,p-1}
\ee
$2p$ of the $4p-2$ ${\cal W}$-indecomposable representations are
projective, namely the two rank-1 
representations $\ketw{\kappa,p}$ and all of the rank-2 representations.
It follows that the two representations $\ketw{\kappa,s}$ are both
${\cal W}$-irreducible and projective.
Since we are only considering the logarithmic minimal models in the ${\cal W}$-extended
picture, we will omit specifications such as {\em ${\cal W}$-irreducible} and simply write
{\em irreducible} in the following.

\subsection{Fusion algebra}

We denote the fusion multiplication in the ${\cal W}$-extended picture by $\!\fus\!$.
The fusion rules underlying the commutative and associative fusion algebra of ${\cal WLM}(1,p)$ 
read~\cite{GK9606,EF0604,GR0707,PRR0803} 
\bea
 \ketw{\kappa,s}\fus\ketw{\kappa',s'}&=&
  \!\bigoplus_{j=|s-s'|+1,\ \!\mathrm{by}\ \!2}^{p-|p-s-s'|-1}
  \!\!\!\ketw{\kappa\cdot\kappa',j}
    \oplus\!\!\bigoplus_{\beta=\eps(s+s'-p-1),\ \!\mathrm{by}\ \!2}^{s+s'-p-1}
    \!\!\!\ketw{\R_{\kappa\cdot\kappa'}^{\beta}}   
   \nn
 \ketw{\kappa,s}\fus\ketw{\R_{\kappa'}^{b}}&=&
    \!\bigoplus_{\beta=|s-b|+1,\ \!\mathrm{by}\ \!2}^{p-|p-s-b|-1}
      \!\!\!\!\ketw{\R_{\kappa\cdot\kappa'}^{\beta}}
     \oplus\!\!\bigoplus_{\beta=\eps(s-b-1),\ \!\mathrm{by}\ \!2}^{s-b-1}
        \!\!\!\!2\ketw{\R_{\kappa\cdot\kappa'}^{\beta}}\oplus
     \!\!\!\bigoplus_{\beta=\eps(s+b-p-1),\ \!\mathrm{by}\ \!2}^{s+b-p-1}
        \!\!\!\!2\ketw{\R_{2\cdot\kappa\cdot\kappa'}^{\beta}}  
    \nn
 \ketw{\R_{\kappa}^{b}}\fus\ketw{\R_{\kappa'}^{b'}}
  &=&\bigoplus_{\beta=\eps(p-b-b'-1),\ \!\mathrm{by}\ \!2}^{p-|b-b'|-1}
     \!\!\!\!2\ketw{\R_{\kappa\cdot\kappa'}^{\beta}}
     \oplus\!\!\bigoplus_{\beta=\eps(p-b-b'-1),\ \!\mathrm{by}\ \!2}^{|p-b-b'|-1}
     \!\!\!2\ketw{\R_{\kappa\cdot\kappa'}^{\beta}}\nn
   &&\qquad\oplus\!\bigoplus_{\beta=\eps(b+b'-1),\ \!\mathrm{by}\ \!2}^{p-|p-b-b'|-1}
     \!\!\!2\ketw{\R_{2\cdot\kappa\cdot\kappa'}^{\beta}}
     \oplus\!\bigoplus_{\beta=\eps(b+b'-1),\ \!\mathrm{by}\ \!2}^{|b-b'|-1}
     \!\!\!2\ketw{\R_{2\cdot\kappa\cdot\kappa'}^{\beta}}  
\label{fus}
\eea
where we have introduced $\ketw{\R_{\kappa}^0}\equiv\ketw{\kappa,p}$ and
\be
 \eps(n)\ =\ \frac{1-(-1)^n}{2},\qquad\quad
   n\cdot m\ =\ 1+\eps(n+m)\ =\ \frac{3-(-1)^{n+m}}{2},\qquad\quad n,m\in\mathbb{Z}
\label{eps}
\ee
We note that this dot product is associative.
The irreducible representation $\ketw{1,1}$ is the fusion-algebra identity,
and the fusion algebra is seen to be generated from repeated fusions of the two fundamental
representations $\ketw{2,1}$ and $\ketw{1,2}$. 
The works~\cite{GK9606,EF0604,GR0707,PRR0803} provide considerable evidence 
for the validity of these fusion rules, though a rigorous proof is not known at present.

\subsection{Fusion matrices and polynomial fusion ring}
\label{SecPolRing}

The fusion algebra, see~\cite{DiFMS} for example,
\be
 \phi_i\otimes\phi_j\ =\ \bigoplus_{k\in\mathcal{J}}{N_{i,j}}^k\phi_k,\hspace{1cm}i,j\in\mathcal{J}
\ee
of a {\em rational} conformal field theory is finite (since the set $\mathcal{J}$ of fusion-algebra
generators is finite) and
can be represented by a commutative matrix algebra $\langle \Nc_i;\ i\in\mathcal{J}\rangle$ 
where the entries of the $|\mathcal{J}|\times|\mathcal{J}|$ matrix $\Nc_i$ are
\be
 {(\Nc_i)_j}^k\ =\ {N_{i,j}}^k,\hspace{1cm}i,j,k\in\mathcal{J}
\ee
and where the fusion multiplication $\otimes$ has been replaced by ordinary matrix multiplication.
In~\cite{Gep91}, Gepner found that every such algebra is isomorphic to a ring
of polynomials in a finite set of variables modulo an ideal defined as the vanishing conditions
of a finite set of polynomials in these variables. 

With respect to some ordering of the fusion generators in (\ref{Jp}), we let
\be
  \big\{\Nc_{\ketw{\kappa,s}},\Nc_{\ketw{\R_\kappa^b}};\  \kappa\in\mathbb{Z}_{1,2},
     s\in\mathbb{Z}_{1,p}, b\in\mathbb{Z}_{1,p-1}\big\}
\label{NN}
\ee 
denote the set of fusion matrices realizing
the fusion algebra (\ref{fus}) of ${\cal WLM}(1,p)$, where $\Nc_{\ketw{\kappa,s}}$ and
$\Nc_{\ketw{\R_\kappa^b}}$ are the matrix realizations of the 
indecomposable representations $\ketw{\kappa,s}$ and 
$\ketw{\R_\kappa^b}$, respectively. These are all $(4p-2)$-dimensional
square matrices, and we are thus dealing with the {\em regular} representation
of the fusion algebra. Special notation is introduced for the two fundamental
fusion matrices
\be
 X\ =\ \Nc_{\ketw{2,1}},\qquad\quad Y\ =\ \Nc_{\ketw{1,2}}
\label{XY}
\ee
Since we are only considering ${\cal WLM}(1,p)$, we have thus abandoned the normalization
convention of~\cite{Ras0812} and will be using the one appearing in (\ref{XY}).

{}From~\cite{Ras0812}, we have the fusion-matrix realization
\bea
 \Nc_{\ketw{\kappa,s}}&=&\mathrm{pol}_{\ketw{\kappa,s}}(X,Y)
    \ =\ X^{\kappa-1}U_{s-1}(\tfrac{Y}{2})
   \nn
 \Nc_{\ketw{\R_\kappa^b}}&=&\mathrm{pol}_{\ketw{\R_\kappa^b}}(X,Y)
   \ =\ 2X^{\kappa-1}T_b(\tfrac{Y}{2})U_{p-1}(\tfrac{Y}{2})
\label{N}
\eea
of the fusion algebra of ${\cal WLM}(1,p)$, where $T_n$ and $U_n$ are Chebyshev 
polynomials of the first and second kind, respectively. Chebyshev polynomials are ubiquitous 
and discussed in~\cite{MH03}, for example, and in the appendix of~\cite{Ras0812}.

It also follows from~\cite{Ras0812} that this fusion algebra is isomorphic
to the polynomial ring $\mathbb{C}[X,Y]$ modulo the ideal 
$(X^2-1,P_p(Y),\tilde{P}_{1,p}(X,Y))$, that is,
\be
 \big\langle\ketw{\kappa,s},\ketw{\R_\kappa^b};\  \kappa\in\mathbb{Z}_{1,2},
     s\in\mathbb{Z}_{1,p}, b\in\mathbb{Z}_{1,p-1}\big\rangle\ \simeq\ 
   \mathbb{C}[X,Y]/\big(X^2-1,P_p(Y),\tilde{P}_{1,p}(X,Y)\big)
\label{iso}
\ee
where
\be
 P_p(Y)\ =\ (Y^2-4)U_{p-1}^3(\tfrac{Y}{2}),\qquad\quad 
  \tilde{P}_{1,p}(X,Y)\ =\ \big(X-T_{p}(\tfrac{Y}{2})\big)U_{p-1}(\tfrac{Y}{2})
\label{PP}
\ee
The polynomial $\tilde{P}_{1,p}(X,Y)$ differs slightly from the polynomial $P_{1,p}(X,Y)$
in~\cite{Ras0812} due to the modified normalization convention in (\ref{XY}).
As demonstrated in Appendix~\ref{AppIso}, we have
\be
 P_p(Y)\ \equiv\ 0\qquad(\mathrm{mod}\ X^2-1,\tilde{P}_{1,p}(X,Y))
\ee
and hence 
\be
 \big\langle\ketw{\kappa,s},\ketw{\R_\kappa^b};\  \kappa\in\mathbb{Z}_{1,2},
     s\in\mathbb{Z}_{1,p}, b\in\mathbb{Z}_{1,p-1}\big\rangle\ \simeq\ 
   \mathbb{C}[X,Y]/\big(X^2-1,\tilde{P}_{1,p}(X,Y)\big)
\label{iso2}
\ee
simplifying the description of the right-hand side of the isomorphism (\ref{iso}).

It is noted that $X$ and $Y$ in (\ref{iso}) and (\ref{iso2}) are formal entities and hence need 
not be identified with the fusion matrices $X$ and $Y$ in (\ref{XY}) and (\ref{N}).
It is nevertheless convenient to use the same notation in the two situations.
Using the explicit fusion matrices and their Jordan decompositions to be discussed below,
the quotient polynomial ring conditions in (\ref{iso}) are verified partly by 
(\ref{charX}) and otherwise in Appendix~\ref{AppQuotient}.

\section{Explicit fusion matrices}

The set of fusion generators (\ref{Jp}) is distinguished in the sense that the associated fusion rules 
(\ref{fus}) involve only non-negative integer multiplicities. It turns out that the ordering
\bea
 &&\ketw{1,1},\ketw{2,1};\ldots;\ketw{1,s},\ketw{2,s};\ldots;\ketw{1,p},\ketw{2,p};\nn
 &&\qquad\qquad\ketw{\R_1^1},\ketw{\R_2^1};\ldots;
   \ketw{\R_1^b},\ketw{\R_2^b};\ldots;\ketw{\R_1^{p-1}},\ketw{\R_2^{p-1}}
\label{order}
\eea
provides a convenient basis in which to study the fusion matrices, 
and is the one used in the following.
It is recalled that we are working with the regular representation of the fusion algebra.

It should be clear that
\be
 \Nc_{\ketw{1,1}}\ =\ I
\ee
The fusion matrices $X$ and $Y$ are given in (\ref{XCY}) below.
The fusion-matrix realizations of the remaining generators in (\ref{Jp}) can all be
obtained by polynomial constructions (\ref{N}) from the realizations $X$ and $Y$
of the fundamental representations $\ketw{2,1}$ and $\ketw{1,2}$.

\subsection{Fundamental fusion matrices}

To facilitate the description of the fundamental fusion matrices, we introduce
\be
 0_2\ =\ \left(\!\!\begin{array}{cc} 0&0 \\ 0&0\end{array}\!\!\right),\qquad\quad
 I_2\ =\ \left(\!\!\begin{array}{cc} 1&0 \\ 0&1\end{array}\!\!\right),\qquad\quad
 C_2\ =\ \left(\!\!\begin{array}{cc} 0&1 \\ 1&0\end{array}\!\!\right)
\ee
For simplicity, we use $I=I_{m}$ and $C=C_{2n}$ to denote the $m$-dimensional identity matrix
and the $2n$-dimensional square matrix
\be
 C\ =\ \mathrm{diag}(\underbrace{C_2,\ldots,C_2}_{n})
\ee
respectively, when their dimensions are understood from the context. We note that $C$ is
an involutory matrix, $C^2=I$.

In the basis (\ref{order}), the fundamental fusion matrices read
\be
 X\ =\ C,\qquad\quad Y\ =\ 
\renewcommand{\arraystretch}{1.2}
\left(\!\!
\begin{array}{cccccc|c|ccccccc}
  0_2&I_2&0_2&\cdots&&&&&&&&& \\
  I_2&0_2&I_2&\ddots&&&&&&&&& \\
  0_2&I_2&0_2&\ddots&&&&&&&&& \\
  \vdots&\ddots&\ddots&\ddots&&&&&&&&& \\
  &&&&0_2&I_2&&&&&&& \\
  &&&&I_2&0_2&I_2&&&&&& \\
\hline
  &&&&&0_2&0_2&I_2&&&&& \\
\hline
  &&&&&&2I_2&0_2&I_2&&&& \\
  &&&&&&&I_2&0_2&I_2&&& \\
  &&&&&&&&I_2&0_2&\ddots&& \\
  &&&&&&&&&\ddots&\ddots& \\
  &&&&&&&&&&&0_2&I_2 \\
  &&&&&&2C_2&&&&&I_2&0_2
\end{array}
\!\!\right)
\label{XCY}
\ee
The $(4p-2)$-dimensional matrix $Y$ is written here as a $(2p-1)$-dimensional matrix, with 
$2\times2$ matrices as entries, whose $p$'th row and column are emphasized to indicate their
special status.
For small values of $p$, the expression (\ref{XCY}) for $Y$ is meant to reduce to
\bea
 Y\big|_{p=2}\ =\ \left(\!\!\begin{array}{ccc} 0_2&I_2&0_2 \\ 0_2&0_2&I_2 \\ 0_2&2I_2+2C_2&0_2
    \end{array}\!\!\right),\qquad\quad
 Y\big|_{p=3}\ =\ \left(\!\!\begin{array}{ccccc} 0_2&I_2&0_2&0_2&0_2 \\ I_2&0_2&I_2&0_2&0_2 \\
    0_2&0_2&0_2&I_2&0_2 \\ 0_2&0_2&2I_2&0_2&I_2 \\ 0_2&0_2&2C_2&I_2&0_2
         \end{array}\!\!\right)
\label{Yp}
\eea
As required, it follows from (\ref{XCY}) that $X$ and $Y$ commute.

The minimal and characteristic polynomials of $X$ are readily seen to be
\be
 X^2-I\ =\ (X-I)(X+I),\qquad\quad  \det(\la I-X)\ =\ (\la-1)^{2p-1}(\la+1)^{2p-1}
\label{charX}
\ee
The description of the similar polynomials for $Y$ uses that
the Chebyshev polynomial $U_{p-1}(x)$ factorizes as
\be
 U_{p-1}(x)\ =\ 2^{p-1}\prod_{j=1}^{p-1}(x-\al_j),\qquad\quad
      \al_j\ =\ \cos\theta_j,\qquad \theta_j\ =\ \frac{j\pi}{p},\qquad j\in\mathbb{Z}_{0,p}
\label{Ua}
\ee
where we have included definitions of $\al_0=1$ and $\al_p=-1$.
In accordance with (\ref{PP}), and as we shall verify explicitly in Appendix~\ref{AppQuotient},
the minimal and characteristic polynomials of $Y$ are
\bea
  P_p(Y)&=&(Y^2-4I)U_{p-1}^3(\tfrac{Y}{2})
    \ =\ (Y-2I)(Y+2I)\prod_{j=1}^{p-1}(Y-2\al_jI)^3\nn
 \det(\la I-Y)&=&U_{p-1}(\tfrac{\la}{2})P_p(\la)\ =\ 
    (\la^2-4)U_{p-1}^4(\tfrac{\la}{2})\ =\ (\la-2)(\la+2)\prod_{j=1}^{p-1}(\la-2\al_j)^4
\label{charY}
\eea
This implies that the Jordan canonical form of $Y$ consists of $p-1$ rank-3 blocks
associated to the eigenvalues $\beta_b=2\al_b$, $b\in\mathbb{Z}_{1,p-1}$, 
and $p+1$ rank-1 blocks associated to the eigenvalues $\beta_j=2\al_j$, $j\in\mathbb{Z}_{0,p}$.
The number of linearly independent eigenvectors of $Y$ is thus $2p$. Since 
the null space of $Y$ is empty for $p$ odd but two-dimensional for $p$ even, the rank of $Y$ is
\be
 \mathrm{rank}(Y)\ =\ 4p-2-2\eps(p-1)\ =\ 4(p-1)+2\eps(p)
\label{rankY}
\ee

\section{Fusion graphs}
\label{SecGraph}

The fusion matrices (\ref{NN}) are mutually commuting, but in general not diagonalizable. 
Nevertheless, we will show that they can be simultaneously brought to Jordan form 
by a similarity transformation, and that the associated similarity matrix 
is determined from the modular data. Prior to demonstrating these important
results, we here discuss the two graphs whose underlying adjacency 
matrices are given by the fundamental fusion matrices $X=\Nc_{\ketw{2,1}}$ and 
$Y=\Nc_{\ketw{1,2}}$. In this context,
\be
 \Nc_\mu \Nc_\nu\ =\ \sum_{\la\in\Ic} \Nc_{\mu,\nu}{}^\la \Nc_\la
\ee
is referred to as the {\em graph fusion algebra}. Here $\Ic$ is the set of indecomposable
representations given in (\ref{NN}). To simplify the notation, we introduce
\be
 N_{\kappa,s}\ =\ \Nc_{\ketw{\kappa,s}},\qquad\quad
 N_\kappa^b\ =\ \Nc_{\ketw{\R_\kappa^b}}
\ee

Fusion graphs succinctly encode the fusion rules and have been instrumental in the 
classification of rational conformal field theories on the cylinder~\cite{BPPZ9809,BPPZ9908} 
and on the torus~\cite{Ocn99,Ocn00,PZ0011,PZ0101}. 
In the rational $A$-type theories, the Verlinde algebra yields a diagonal modular
invariant, while $D$- and $E$-type theories are related to non-diagonal
modular invariants. The Ocneanu algebras arise when considering fusion on the
torus, with left and right chiral halves of the theory, and involve Ocneanu graphs.
We refer to~\cite{Kos88,DiFZ90,DiF92,PZ9510} for earlier results on the interrelation
between fusion algebras, graphs and modular invariants.

The fundamental fusion graph associated to $Y$ follows from (\ref{XCY}).
For $p=4$, in particular, it is given by
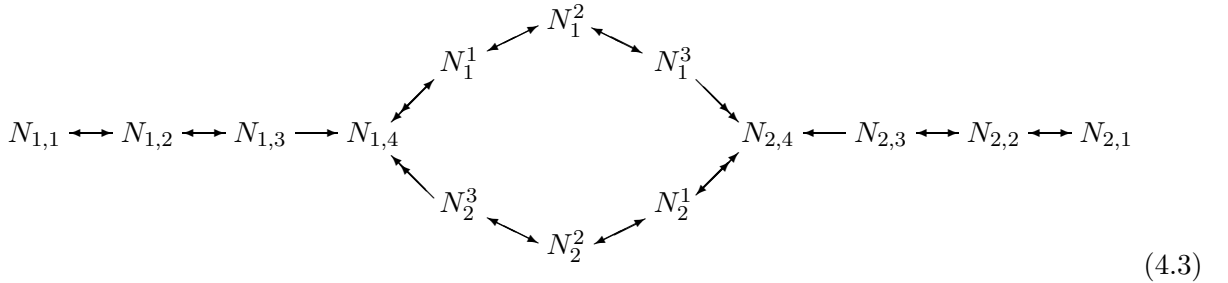
\be
 \mbox{
 \begin{picture}(100,110)(160,-50)
    \unitlength=0.75cm
  \thinlines
\put(0,0){$N_{1,1}$}
\put(1.5,0.15){\vector(-1,0){0.4}}
\put(1.5,0.15){\vector(1,0){0.35}}
\put(2,0){$N_{1,2}$}
\put(3.5,0.15){\vector(-1,0){0.4}}
\put(3.5,0.15){\vector(1,0){0.35}}
\put(4,0){$N_{1,3}$}
\put(5.1,0.15){\vector(1,0){0.75}}
\put(6,0){$N_{1,4}$}
\put(7.45,1){\vector(-1,-1){0.5}}
\put(7.3,0.85){\vector(-1,-1){0.5}}
\put(7.05,0.6){\vector(1,1){0.5}}
\put(7.65,1.25){$N_1^1$}
\put(8.8,1.75){\vector(2,1){0.6}}
\put(9,1.85){\vector(-2,-1){0.5}}
\put(9.55,2){$N_1^2$}
\put(10.7,1.85){\vector(2,-1){0.55}}
\put(10.9,1.75){\vector(-2,1){0.55}}
\put(11.45,1.25){$N_1^3$}
\put(12.2,1.1){\vector(1,-1){0.6}}
\put(12.35,0.95){\vector(1,-1){0.6}}
\put(7.55,-1){\vector(-1,1){0.6}}
\put(7.4,-0.85){\vector(-1,1){0.6}}
\put(7.65,-1.25){$N_2^3$}
\put(8.8,-1.5){\vector(2,-1){0.6}}
\put(9,-1.6){\vector(-2,1){0.5}}
\put(9.55,-2){$N_2^2$}
\put(10.95,-1.55){\vector(-2,-1){0.55}}
\put(10.75,-1.65){\vector(2,1){0.55}}
\put(11.45,-1.25){$N_2^1$}
\put(12.3,-0.85){\vector(1,1){0.5}}
\put(12.45,-0.7){\vector(1,1){0.5}}
\put(12.7,-0.45){\vector(-1,-1){0.5}}
\put(13,0){$N_{2,4}$}
\put(14.85,0.15){\vector(-1,0){0.75}}
\put(15,0){$N_{2,3}$}
\put(16.5,0.15){\vector(-1,0){0.4}}
\put(16.5,0.15){\vector(1,0){0.35}}
\put(17,0){$N_{2,2}$}
\put(18.5,0.15){\vector(-1,0){0.4}}
\put(18.5,0.15){\vector(1,0){0.35}}
\put(19,0){$N_{2,1}$}
 \end{picture}
}
\label{gYp4}
\ee
This is readily extended to general $p$ where we have
\be
 \mbox{
 \begin{picture}(100,100)(160,-45)
    \unitlength=0.75cm
  \thinlines
\put(-0.5,0){$N_{1,1}$}
\put(1,0.15){\vector(-1,0){0.4}}
\put(1,0.15){\vector(1,0){0.35}}
\put(1.7,0){$\ldots$}
\put(3,0.15){\vector(-1,0){0.4}}
\put(3,0.15){\vector(1,0){0.35}}
\put(3.5,0){$N_{1,p-1}$}
\put(5,0.15){\vector(1,0){0.85}}
\put(6,0){$N_{1,p}$}
\put(7.45,1){\vector(-1,-1){0.5}}
\put(7.3,0.85){\vector(-1,-1){0.5}}
\put(7.05,0.6){\vector(1,1){0.5}}
\put(7.65,1.25){$N_1^1$}
\put(8.8,1.75){\vector(2,1){0.6}}
\put(9,1.85){\vector(-2,-1){0.5}}
\put(9.55,2){$\ldots$}
\put(10.7,1.85){\vector(2,-1){0.55}}
\put(10.9,1.75){\vector(-2,1){0.55}}
\put(11.45,1.25){$N_1^{p-1}$}
\put(12.2,1.1){\vector(1,-1){0.6}}
\put(12.35,0.95){\vector(1,-1){0.6}}
\put(7.55,-1){\vector(-1,1){0.6}}
\put(7.4,-0.85){\vector(-1,1){0.6}}
\put(7.65,-1.25){$N_2^{p-1}$}
\put(8.8,-1.55){\vector(2,-1){0.6}}
\put(9,-1.65){\vector(-2,1){0.5}}
\put(9.6,-1.9){$\ldots$}
\put(10.95,-1.6){\vector(-2,-1){0.55}}
\put(10.75,-1.7){\vector(2,1){0.55}}
\put(11.45,-1.25){$N_2^1$}
\put(12.3,-0.85){\vector(1,1){0.5}}
\put(12.45,-0.7){\vector(1,1){0.5}}
\put(12.7,-0.45){\vector(-1,-1){0.5}}
\put(13,0){$N_{2,p}$}
\put(14.85,0.15){\vector(-1,0){0.75}}
\put(15,0){$N_{2,p-1}$}
\put(17,0.15){\vector(-1,0){0.4}}
\put(17,0.15){\vector(1,0){0.35}}
\put(17.65,0){$\ldots$}
\put(19,0.15){\vector(-1,0){0.4}}
\put(19,0.15){\vector(1,0){0.35}}
\put(19.5,0){$N_{2,1}$}
 \end{picture}
}
\label{gYp}
\ee
which, for $p=2$, reduces to
\be
 \mbox{
 \begin{picture}(100,80)(120,-30)
    \unitlength=0.75cm
  \thinlines
\put(4,0){$N_{1,1}$}
\put(5.1,0.15){\vector(1,0){0.75}}
\put(6,0){$N_{1,2}$}
\put(7.45,1){\vector(-1,-1){0.5}}
\put(7.3,0.85){\vector(-1,-1){0.5}}
\put(7.05,0.6){\vector(1,1){0.5}}
\put(7.65,1.25){$N_1^1$}
\put(8.4,1.1){\vector(1,-1){0.6}}
\put(8.55,0.95){\vector(1,-1){0.6}}
\put(7.55,-1){\vector(-1,1){0.6}}
\put(7.4,-0.85){\vector(-1,1){0.6}}
\put(7.65,-1.25){$N_2^1$}
\put(8.5,-0.85){\vector(1,1){0.5}}
\put(8.65,-0.7){\vector(1,1){0.5}}
\put(8.9,-0.45){\vector(-1,-1){0.5}}
\put(9.2,0){$N_{2,1}$}
\put(11.05,0.15){\vector(-1,0){0.75}}
\put(11.2,0){$N_{2,2}$}
 \end{picture}
}
\label{gYp2}
\ee
The fundamental fusion graph associated to $X$ has $2p-1$ disconnected components
\be
 \mbox{
 \begin{picture}(100,20)(120,0)
    \unitlength=0.75cm
  \thinlines
\put(4,0){$N_{1,s}$}
\put(5.5,0.15){\vector(1,0){0.35}}
\put(5.5,0.15){\vector(-1,0){0.4}}
\put(6,0){$N_{2,s}$}
\put(9.2,0){$N_1^b$}
\put(10.6,0.15){\vector(-1,0){0.45}}
\put(10.6,0.15){\vector(1,0){0.4}}
\put(11.2,0){$N_2^b$}
 \end{picture}
}
\ee
where it is recalled that $s\in\mathbb{Z}_{1,p}$ and $b\in\mathbb{Z}_{1,p-1}$.
The fundamental fusion graph associated to $Y$ is covariant under the action of 
$X$ in the sense that $X$ acts by rotating
the graph, as it is depicted in (\ref{gYp}), by $180^{\circ}$. Ignoring the labeling of the $N$'s
appearing in (\ref{gYp}), the graph itself is invariant under rotation by $180^{\circ}$.

It is recalled that there are $2p$ irreducible representations, $\ketw{\kappa,s}$, and
$2p$ projective representations, $\ketw{\kappa,p}$ and $\ketw{\R_\kappa^b}$,
where the two representations $\ketw{\kappa,p}$ are both irreducible and projective.
The two horizontal legs of the fusion graph (\ref{gYp}) are composed of the
irreducible representations, while the loop consists of the projective representations
with $N_{\kappa,p}\sim\ketw{\kappa,p}$ appearing in both a horizontal leg and the loop.
Combined with the two one-way arrows linking $N_{\kappa,p-1}\to N_{\kappa,p}$,
this reflects that the set of projective representations forms an ideal of the fusion algebra
(\ref{fus}).

\section{Spectral decompositions}
\label{SecSpectral}

In preparation for the spectral decomposition of the various fusion matrices,
we recall that the canonical rank-3 Jordan block associated to the eigenvalue $\la$ is given by
\be
   \Jc_{\la,3}\ =\ \left(\!\!\begin{array}{ccc} \la&1&0 \\ 0&\la&1 \\ 0&0&\la \end{array}\!\!\right)
\label{Jb}
\ee 
It is sometimes convenient to relax the condition of unity in the super-diagonal of a
Jordan block. We thus refer to any matrix of the form
\be
 \left(\!\!\begin{array}{ccc} \la&\nu_1&\nu_3 \\ 0&\la&\nu_2 \\ 0&0&\la \end{array}\!\!\right),\qquad
 \quad \nu_1\neq0,\ \nu_2\neq0
\ee
as a Jordan block of rank 3 associated to the eigenvalue $\la$.
The Jordan {\em canonical} block $\Jc_{\la,3}$ is recovered by setting
$\nu_1,\nu_2=1$ and $\nu_3=0$. A block-diagonal matrix consisting of (canonical)
Jordan blocks only is said to be in Jordan (canonical) form. 
Also, for a function $g$ expandable as a power series, we note that
\be
 g\big(\!\left(\!\!\begin{array}{ccc} \la&\nu_1&\nu_3 \\ 0&\la&\nu_2 
  \\ 0&0&\la \end{array}\!\!\right)\!\big)
   \ =\ \left(\!\!\begin{array}{ccc} g(\la)&\nu_1g'(\la)&\nu_3g'(\la)+\hf\nu_1\nu_2g''(\la) 
     \\ 0&g(\la)&\nu_2g'(\la) \\ 0&0&g(\la) \end{array}\!\!\right)
\label{g}
\ee

Our first objective in this section is to devise a similarity transformation in the form of a 
matrix $Q$ which Jordan-decomposes $X$ and $Y$ simultaneously
\be
 Q^{-1}XQ\ =\ J_X,\qquad\quad Q^{-1}YQ\ =\ J_Y
\label{QXQ}
\ee
where $J_X$ and $J_Y$ are Jordan canonical forms.
For every $\Nc$ in (\ref{NN}), it then follows that
\be
 Q^{-1}\Nc Q\ =\ Q^{-1}\mathrm{pol}_{\Nc}(X,Y)Q\ =\ \mathrm{pol}_{\Nc}(Q^{-1}XQ,Q^{-1}YQ)\ =\ 
   \mathrm{pol}_{\Nc}(J_X,J_Y)
\label{QNQ}
\ee
implying that $Q$ also brings $\Nc$ to Jordan form, albeit not necessarily Jordan
{\em canonical} form. Our second objective is therefore
to find invertible matrices $\hat{Q}_\Nc$ such that
\be
 Q_{\Nc}\ =\ Q\hat{Q}_{\Nc}
\label{QQ}
\ee
coverts $\Nc$ into
\be
 Q^{-1}_{\Nc}\Nc Q_{\Nc}\ =\ \hat{Q}^{-1}_{\Nc}\mathrm{pol}_{\Nc}(J_X,J_Y)\hat{Q}_{\Nc}\ =\ J_{\Nc}
\label{QNQJ}
\ee
where $J_{\Nc}$ {\em is} a Jordan canonical form.

\subsection{Fundamental fusion matrices}

For $k\in\mathbb{Z}_{1,2p-1}$, we define the function $f_k(x)$ by
\be
 f_s(x)\ =\ U_{s-1}(\tfrac{x}{2}),\qquad\quad 
 f_{p+b}(x)\ =\ 2T_b(\tfrac{x}{2})U_{p-1}(\tfrac{x}{2})\ =\ U_{p+b-1}(\tfrac{x}{2})
    +U_{p-b-1}(\tfrac{x}{2})
\label{f}
\ee
where it is recalled that $s\in\mathbb{Z}_{1,p}$ and $b\in\mathbb{Z}_{1,p-1}$.
These functions describe the $Y$-parts of (\ref{N}) 
\be
 \Nc_{\ketw{\kappa,s}}\ =\ X^{\kappa-1}f_s(Y),\qquad\quad
 \Nc_{\ketw{\R_\kappa^b}}\ =\ X^{\kappa-1}f_{p+b}(Y)
\label{NXf}
\ee
and are used below in the construction of the similarity matrix $Q$.
Certain properties of $f_k(x)$ are described in Appendix~\ref{AppPropfk}.
In the following, we will initially assume that $p>2$ and subsequently consider the case $p=2$.

\subsubsection{$p>2$}

Let us introduce the $p+1$ $(4p-2)$-dimensional vectors
\be
 v_0\ =\ \left(\!\!\begin{array}{c} f_1(\beta_0)\vb_0 \\ \vdots \\ f_{2p-1}(\beta_0)\vb_0 
    \end{array}\!\!\right), \qquad\quad
 v_b\ =\ \left(\!\!\begin{array}{c} f_1(\beta_b)\vb_{b-1} \\ \vdots \\ f_{2p-1}(\beta_b)\vb_{b-1} 
    \end{array}\!\!\right), \qquad\quad     
 v_p\ =\ \left(\!\!\begin{array}{c} f_1(\beta_p)\vb_p \\ \vdots \\ f_{2p-1}(\beta_p)\vb_p \end{array}\!\!\right)
\label{0p}
\ee
where we draw attention to the different conventions for the indices of the auxiliary 
two-dimensional vectors
\be
 \vb_n\ =\ \left(\!\!\begin{array}{c} 1 \\ (-1)^n \end{array}\!\!\right),\qquad\quad n\in\mathbb{Z}
\label{v}
\ee
For every $b\in\mathbb{Z}_{1,p-1}$, we also introduce the triplet of $(4p-2)$-dimensional
vectors
\be
 w_b^{(1)}\ =\ \left(\!\!\begin{array}{c} f_1(\beta_b)\vb_{b} \\ \vdots \\ f_{2p-1}(\beta_b)\vb_{b} 
    \end{array}\!\!\right), \qquad\quad     
 w_b^{(2)}\ =\ \left(\!\!\begin{array}{c} f'_1(\beta_b)\vb_{b} \\ \vdots \\ 
   f'_{2p-1}(\beta_b)\vb_{b} 
    \end{array}\!\!\right),\qquad\quad
    w_b^{(3)}\ =\ \left(\!\!\begin{array}{c} \frac{1}{2}f''_1(\beta_b)\vb_{b} 
    \\ \vdots \\  \frac{1}{2}f''_{2p-1}(\beta_b)\vb_{b} 
    \end{array}\!\!\right)
\label{w}
\ee
Using (\ref{ff}), (\ref{fdiff}) and (\ref{f2p}), in particular, it is straightforward to verify that the $p+1$
vectors in (\ref{0p}) are eigenvectors of $Y$ corresponding to the eigenvalues 
$\beta_0$, $\beta_b$ and $\beta_p$, and that, for every $b\in\mathbb{Z}_{1,p-1}$, 
the three vectors in (\ref{w}) form a Jordan chain
\be
 Yw_b^{(3)}\ =\ \beta_b w_b^{(3)}+w_b^{(2)},\qquad \quad
 Yw_b^{(2)}\ =\ \beta_b w_b^{(2)}+w_b^{(1)},\qquad \quad
 Yw_b^{(1)}\ =\ \beta_b w_b^{(1)}
\label{Yw}
\ee
corresponding to the eigenvalue $\beta_b$. This chain of relations imply that
\be
 (Y-\beta_bI) w_b^{(3)}\ =\ w_b^{(2)},\qquad\quad
 (Y-\beta_bI) w_b^{(2)}\ =\ w_b^{(1)},\qquad\quad   
 (Y-\beta_bI)^\ell w_b^{(\ell)}\ =\ 0,\quad \ell\in\mathbb{Z}_{1,3}
\label{YIw}
\ee
where the vanishing conditions indicate that the vectors are generalized eigenvectors.

The two eigenvectors $v_b$ and $w_b^{(1)}$ correspond to the same eigenvalue $\beta_b$ but
are obviously linearly independent. Since $\beta_i\neq\beta_j$ for $i\neq j$, 
the $(4p-2)$-dimensional matrix $Q$ is constructed by concatenating the generalized 
(of which $2p$ are proper) eigenvectors (\ref{0p}) and (\ref{w})
\be
 Q\ =\ \left(\!\!\begin{array}{c|cccc|c|cccc|c} v_0&v_1&w_1^{(1)}&w_1^{(2)}&w_1^{(3)}&\ldots&
    v_{p-1}&w_{p-1}^{(1)}&w_{p-1}^{(2)}&w_{p-1}^{(3)}&v_p  \end{array}\!\!\right)
\label{Qvw}
\ee
By the similarity transformation (\ref{QXQ}), this matrix $Q$ converts $Y$ into its 
Jordan canonical form
\be
 J_Y\ =\ \mathrm{diag}\big(\beta_0;\beta_1,\Jc_1;\ldots; \beta_b,\Jc_b;\ldots; 
     \beta_{p-1},\Jc_{p-1};\beta_p\big),\qquad\quad \Jc_b\ =\ \Jc_{\beta_b,3}
\label{JY}
\ee
where $\Jc_{\beta_b,3}$ is the canonical rank-3 Jordan block (\ref{Jb}) associated to 
the eigenvalue $\beta_b$. Thus, the eigenvalues $\beta_0=2$ and $\beta_p=-2$ both 
have geometric and algebraic multiplicity 1,
whereas the $p-1$ eigenvalues $\beta_b=2\cos\theta_b$ all have geometric multiplicity 2
and algebraic multiplicity 4. This is in accordance with the minimal and characteristic
polynomials of $Y$ in (\ref{charY}).

It is readily verified that the Jordan canonical form (\ref{QXQ}) of $X$ with respect to $Q$ 
in (\ref{Qvw}) is the diagonal matrix
\be
 J_X\ =\ \mathrm{diag}\big(1;1,-I_3;\ldots;(-1)^{b-1},(-1)^bI_3;\ldots;
   (-1)^{p-2},(-1)^{p-1}I_3;(-1)^p\big)
\label{JX}
\ee
The eigenvalues $1$ and $-1$ both appear with geometric and algebraic multiplicity $2p-1$ 
in accordance with the minimal and characteristic polynomials in (\ref{charX}).

In conclusion, the matrix $Q$ (\ref{Qvw}) converts the fundamental 
fusion matrices $X$ and $Y$ simultaneously into their Jordan canonical forms $J_X$ and $J_Y$. 
We note that $J_X$ and $J_Y$ commute since $X$ and $Y$ commute.
Their commutativity also follows directly from their compatible block structures.

As an aside, based on explicit evaluations of the determinant of $Q$ for small values of $p$,
we conjecture that, for general $p$, it is given by
\be
 \mathrm{det}Q\ =\ 32(-1)^p (2p)^{5p-9}
\label{detQ}
\ee

\subsubsection{$p=2$}

For $p=2$, $Y$ is given in (\ref{Yp}) and the expressions (\ref{0p}) and (\ref{w}) yield
the six generalized eigenvectors
\be
 v_0\!\ =\ \!\left(\!\!\begin{array}{c} 1 \\ 1 \\ 2 \\ 2\\ 4\\ 4 
    \end{array}\!\!\right), \
 v_1\!\ =\ \!\left(\!\!\begin{array}{c} 1 \\ 1 \\ 0 \\ 0\\ 0\\ 0 
    \end{array}\!\!\right), \
 w_1^{(1)}\!\ =\ \!\left(\!\!\!\begin{array}{c} 1 \\ -1 \\ 0 \\ 0\\ 0\\ 0 
    \end{array}\!\!\!\right), \
 w_1^{(2)}\!\ =\ \!\left(\!\!\!\begin{array}{c} 0 \\ 0 \\ 1 \\ -1\\ 0\\ 0 
    \end{array}\!\!\!\right), \
 w_1^{(3)}\!\ =\ \!\left(\!\!\!\begin{array}{c} 0 \\ 0 \\ 0 \\ 0\\ 1\\ -1 
    \end{array}\!\!\!\right), \
 v_2\!\ =\ \!\left(\!\!\!\begin{array}{c} 1 \\ 1 \\ -2 \\ -2\\ 4\\ 4 
    \end{array}\!\!\!\right)
\ee
corresponding to the eigenvalues $\beta_0=-2$, $\beta_1=0$ and $\beta_2=2$.
The associated similarity matrix (\ref{Qvw}) reads
\be
 Q\ =\ \left(\!\!\begin{array}{cccccc} 1&1&1&0&0&1 \\   1&1&-1&0&0&1\\   2&0&0&1&0&-2\\
    2&0&0&-1&0&-2\\   4&0&0&0&1&4\\   4&0&0&0&-1&4   
   \end{array}\!\!\right)
\ee
has determinant $\mathrm{det}(Q)=128$ (in accordance with (\ref{detQ})), 
and Jordan-decomposes $X$ and $Y$ simultaneously
\be
 Q^{-1}XQ\ =\ \mathrm{diag}\big(1,1,-1,-1,-1,1\big),\qquad\qquad
 Q^{-1}YQ\ =\ \mathrm{diag}\big(2,0,\Jc_1,-2\big) 
\ee
where the diagonal elements of the canonical rank-3 Jordan block $\Jc_1$ are $\beta_1=0$.

\subsection{General fusion matrices}

The similarity matrix $Q$ brings all fusion matrices $\Nc$ simultaneously
to Jordan form (\ref{QNQ}). Except for the two fundamental fusion matrices $X$ and $Y$,
these Jordan forms are typically {\em non-canonical}. The objective here is to determine
the refinement $Q_\Nc$ (\ref{QQ}) of $Q$ converting the fusion matrix $\Nc$ into the Jordan 
{\em canonical} form (\ref{QNQJ}).
Thus, continuing the Jordan decomposition of $\Nc$ in (\ref{QNQ}) and (\ref{QNQJ}), we have
\bea
 Q_{\Nc}^{-1}\Nc Q_{\Nc}&=&\hat{Q}_{\Nc}^{-1}J_X^{\kappa-1}f(J_Y)\hat{Q}_{\Nc}\\
  &=&\hat{Q}_{\Nc}^{-1}\mathrm{diag}\big(f(\beta_0);\ldots;
    (-1)^{(\kappa-1)(b-1)}f(\beta_b),(-1)^{(\kappa-1)b}f(\Jc_b);\ldots;
   (-1)^{(\kappa-1)p}f(\beta_b)\big)\hat{Q}_{\Nc}  \nonumber
\label{QNQkappa}
\eea
where $b$ runs from 1 to $p-1$, 
while $f=f_k$, $k\in\mathbb{Z}_{1,2p-1}$, is the function partaking in the description of the given 
fusion matrix $\Nc$ (\ref{NXf}). With (\ref{g}) in mind, 
we here list the Jordan decompositions of all three-dimensional upper-triangular
matrices whose entries of a given (super-)diagonal are identical 
\bea
 \left(\!\!\begin{array}{ccc} 1&& \\   &1& \\     &&1
    \end{array}\!\!\!\right)^{\!\!\!-1}\!\!
 \left(\!\!\begin{array}{ccc} \mathrm{a}&& \\   &\mathrm{a}& \\     &&\mathrm{a}
    \end{array}\!\!\!\right)
 \left(\!\!\begin{array}{ccc} 1&& \\   &1& \\     &&1
    \end{array}\!\!\!\right)   
 &=&  \left(\!\!\begin{array}{ccc} \mathrm{a}&& \\   &\mathrm{a}& \\     &&\mathrm{a}
    \end{array}\!\!\!\right)
 \nn
 \left(\!\!\begin{array}{ccc} 0&1&0 \\   1&0&0 \\     0&0&\frac{1}{\mathrm{c}}
    \end{array}\!\!\!\right)^{\!\!\!-1}\!\!
 \left(\!\!\begin{array}{ccc} \mathrm{a}&0&\mathrm{c} \\   &\mathrm{a}&0 \\     &&\mathrm{a}
    \end{array}\!\!\!\right)
 \left(\!\!\begin{array}{ccc} 0&1&0 \\   1&0&0 \\     0&0&\frac{1}{\mathrm{c}}
    \end{array}\!\!\!\right)   
 &=&  \left(\!\!\begin{array}{ccc} \mathrm{a}&0&0 \\   &\mathrm{a}&1 \\     &&\mathrm{a}
    \end{array}\!\!\!\right),\qquad\quad \mathrm{c}\neq0
 \nn
 \left(\!\!\begin{array}{ccc} 1&\frac{\mathrm{c}}{\mathrm{b}^2}&0 \\[3pt]   
     &\frac{1}{\mathrm{b}}&0 \\[2pt]     &&\frac{1}{\mathrm{b}^2}
    \end{array}\!\!\!\right)^{\!\!\!-1}\!\!
 \left(\!\!\begin{array}{ccc} \mathrm{a}&\mathrm{b}&\mathrm{c} \\ &\mathrm{a}&\mathrm{b} \\ 
   &&\mathrm{a}
    \end{array}\!\!\!\right)
 \left(\!\!\begin{array}{ccc} 1&\frac{\mathrm{c}}{\mathrm{b}^2}&0 \\[3pt] 
    &\frac{1}{\mathrm{b}}&0\\[2pt]  &&\frac{1}{\mathrm{b}^2}
    \end{array}\!\!\!\right)   
 &=&  \left(\!\!\begin{array}{ccc} \mathrm{a}&1&0 \\   &\mathrm{a}&1 \\     &&\mathrm{a}
    \end{array}\!\!\!\right),\qquad\quad \mathrm{b}\neq0
\eea
To complete the spectral decomposition of the fusion matrix $\Nc$, by finding the associated
Jordan canonical form $J_{\Nc}$ and similarity matrix $\hat{Q}_{\Nc}$,
it is therefore necessary to determine whether $f_k'(\beta_b)$ or $f_k''(\beta_b)$ is zero
for $k\in\mathbb{Z}_{1,2p-1}$ and $b\in\mathbb{Z}_{1,p-1}$. These possibilities are
classified in Appendix~\ref{AppZeros}.

Now, the results (\ref{12}) immediately confirm that
\be
 J_{\ketw{1,1}}\ =\ I,\qquad J_{\ketw{2,1}}\ =\ J_X,\qquad J_{\ketw{1,2}}\ =\ J_Y
\label{JJJ}
\ee
and
\be
 Q_{\ketw{1,1}}\ =\ Q_{\ketw{2,1}}\ =\ Q_{\ketw{1,2}}\ =\ Q,\qquad\quad
   \hat{Q}_{\ketw{1,1}}\ =\ \hat{Q}_{\ketw{2,1}}\ =\ \hat{Q}_{\ketw{1,2}}\ =\ I
\label{QQQ}
\ee
where we have introduced the simplified notation $J_{\ketw{\kappa,s}}=J_{\Nc_{\ketw{\kappa,s}}}$
and similarly for $Q_{\Nc}$ and $\hat{Q}_{\Nc}$. 
Below, we will also use $J_{\ketw{\R_\kappa^b}}=J_{\Nc_{\ketw{\R_\kappa^b}}}$ and similarly for
$Q_{\Nc}$ and $\hat{Q}_{\Nc}$.
The Jordan canonical forms $J_{\Nc}$ and similarity matrices $\hat{Q}_{\Nc}$ 
of the remaining (cf. (\ref{JJJ}) and (\ref{QQQ}))
fusion matrices $\Nc_{\ketw{\kappa,s}}$ and $\Nc_{\ketw{\R_\kappa^{b'}}}$ 
in (\ref{NN}) depend on the relations between $p$, the labels $\kappa$, $s$ and $b'$,
and the labeling $b$ of the eigenvalues $\beta_b$. In the following, we list these results
for $J_{\Nc}$ and $\hat{Q}_{\Nc}$, recalling that the
similarity matrix Jordan-decomposing $\Nc$ is given by $Q_\Nc=Q\hat{Q}_\Nc$ (\ref{QQ}).
To this end, we introduce
\be
 \Qc_{g(x)}\ =\  \left(\!\!\begin{array}{ccc} 1&\frac{g'(x)}{2g^2(x)}&0 \\   0&\frac{1}{g(x)}&0 \\     
   0&0&\frac{1}{g^2(x)}  \end{array}\!\!\!\right),\qquad\qquad
 \Jc_{\la,3}^{(1,2)}\ =\  \left(\!\!\begin{array}{ccc} \la&0&0 \\   0&\la&1 \\     
   0&0&\la  \end{array}\!\!\!\right)
\ee
where $g$ is a polynomial evaluated at a point where $g(x)\neq0$. 

Now, for $p$ odd, we have $f_k'(\beta_b)\neq0$ for $k\in\mathbb{Z}_{2,2p-1}$, implying
that, for $s\in\mathbb{Z}_{2,p}$, 
\bea
 J_{\ketw{\kappa,s}}&=&\mathrm{diag}\Big(s;\ldots;(-1)^{(\kappa-1)(b-1)}f_s(\beta_b),
   \Jc_{(-1)^{(\kappa-1)b}f_s(\beta_b),3}; \ldots; s(-1)^{\kappa+s}\Big)   \nn
 J_{\ketw{\kappa,p}}&=&\mathrm{diag}\Big(p;\ldots;0,\Jc_{0,3}; \ldots; p(-1)^{\kappa+1}\Big)   \nn
 J_{\ketw{\R_\kappa^{b'}}}&=&\mathrm{diag}\Big(2p;\ldots;0,
    \Jc_{0,3};\ldots; 2p(-1)^{\kappa+b'-1}\Big)   \nn
 \hat{Q}_{\Nc}&=&\mathrm{diag}\Big(1;\ldots;1,
    \Qc_{(-1)^{(\kappa-1)b}f'(\beta_b)};\ldots;1\Big)
\eea
where $f=f_s$ for $\Nc=\Nc_{\ketw{\kappa,s}}$ while $f=f_{p+b'}$ for 
$\Nc=\Nc_{\ketw{\R_\kappa^{b'}}}$.
For $p$ even, we let $b_1\in\mathbb{Z}_{1,\frac{p}{2}-1}$ and 
$b_2\in\mathbb{Z}_{\frac{p}{2}+1,p-1}$ and find
\bea
 J_{\ketw{\kappa,s}}&=&\mathrm{diag}\Big(s;\ldots;(-1)^{(\kappa-1)(b_1-1)}f_s(\beta_{b_1}),
   \Jc_{(-1)^{(\kappa-1)b_1}f_s(\beta_{b_1}),3};\ldots;0,\Jc_{0,3};\ldots;\nn
  &&\qquad\qquad\qquad\qquad (-1)^{(\kappa-1)(b_2-1)}f_s(\beta_{b_2}),
   \Jc_{(-1)^{(\kappa-1)b_2} f_s(\beta_{b_2}),3};\ldots;-s      \Big)    \nn
  J_{\ketw{\kappa,p}}&=&\mathrm{diag}\Big(p;\ldots;0,\Jc_{0,3};\ldots;-p \Big)  \nn
 \hat{Q}_{\ketw{\kappa,s}}&=&\mathrm{diag}\Big(1;\ldots;1,\Qc_{(-1)^{(\kappa-1)b_1}
    f_s'(\beta_{b_1})};\ldots;1,1,\frac{(-1)^{(\kappa-1)\frac{p}{2}+j-1}}{j},\frac{1}{j^2};\ldots;\nn
   &&\qquad\qquad\qquad\qquad 1,\Qc_{(-1)^{(\kappa-1)b_2}f_s'(\beta_{b_2})};\ldots;1      \Big)   
\eea
for $s=2j$, $j\in\mathbb{Z}_{2,\frac{p}{2}}$, and
\bea
 J_{\ketw{\kappa,s}}&=&\mathrm{diag}\Big(s;\ldots;(-1)^{(\kappa-1)(b_1-1)}f_s(\beta_{b_1}),
    \Jc_{(-1)^{(\kappa-1)b_1}f_s(\beta_{b_1}),3}; \ldots;
     (-1)^{(\kappa-1)(\frac{p}{2}-1)+j},\nn
 &&\qquad  \quad    \Jc_{(-1)^{(\kappa-1)\frac{p}{2}+j},3}^{(1,2)}; \ldots;
    (-1)^{(\kappa-1)(b_2-1)}f_s(\beta_{b_2}),\Jc_{(-1)^{(\kappa-1)b_2}f_s(\beta_{b_2}),3}
    ;\ldots; s\Big)\nn
  \hat{Q}_{\ketw{\kappa,s}}&=& \mathrm{diag}\Big(  1;\ldots;1,\Qc_{(-1)^{(\kappa-1)b_1}
    f_s'(\beta_{b_1})};\ldots;1,C_2,\frac{2(-1)^{(\kappa-1)\frac{p}{2}+j-1}}{j(j+1)};\ldots;\nn
  &&\qquad\qquad\qquad\qquad
   1,\Qc_{(-1)^{(\kappa-1)b_2}f_s'(\beta_{b_2})};\ldots;1    \Big) 
\eea
for $s=2j+1$, $j\in\mathbb{Z}_{1,\frac{p}{2}-1}$.
For $\frac{p}{2(b',p)}\notin\mathbb{N}$, $b'$ is necessarily even and we have
\bea
 J_{\ketw{\R_\kappa^{b'}}}&=&\mathrm{diag}\Big(2p;\ldots;0,\Jc_{0,3};\ldots;-2p\Big)\nn
 \hat{Q}_{\ketw{\R_\kappa^{b'}}}&=&\mathrm{diag}\Big(1;\ldots;1,\Qc_{(-1)^{(\kappa-1)b_1}
    f_{p+b'}'(\beta_{b_1})};\ldots;1,1,\frac{(-1)^{\frac{\kappa p+b'}{2}-1}}{p},\frac{1}{p^2};\ldots;\nn
   &&\qquad\qquad\qquad\quad 1,\Qc_{(-1)^{(\kappa -1)b_2}f_{p+b'}'(\beta_{b_2})};\ldots;1      \Big)
\eea
whereas for $\frac{p}{2(b',p)}\in\mathbb{N}$, we have
\bea
 J_{\ketw{\R_\kappa^{b'}}}&=&\mathrm{diag}\Big(2p;\ldots;0,J_{b',b};\ldots;(-1)^{b'}2p\Big)\nn
 \hat{Q}_{\ketw{\R_\kappa^{b'}}}&=&\mathrm{diag}\Big(1;\ldots;1,Q_{b',b};\ldots;1      \Big)
\eea
where
\be
 \begin{array}{llll}
 &J_{b',b}\ =\ \Jc_{0,3}^{(1,2)},\qquad &Q_{b',b}\ =\ \mathrm{diag}\Big(
    C_2,\frac{2(-1)^{(\kappa-1)b}}{f_{p+b'}''(\beta_b)}\Big),
  \qquad&b=\frac{(2j-1)p}{2(b',p)},\quad j\in\mathbb{Z}_{1,(b',p)} \\[3pt]
 &J_{b',b}\ =\ \Jc_{0,3},\qquad &Q_{b',b}\ =\ \Qc_{(-1)^{(\kappa-1)b}f_{p+b'}'(\beta_b)},
   &\qquad\qquad \mathrm{otherwise}   \end{array}
\ee
Further simplifications are possible but not included here.
Above, $(n,m)$ denotes the greatest common divisor of the integers $n$ and $m$.

\section{Generalized Jordan chains and non-canonical Jordan forms}

In preparation for the description of the Jordan decompositions of the fusion matrices 
in terms of modular data in Section~\ref{SecGenVerFor}, we here present a particularly
convenient similarity matrix which brings all the fusion matrices simultaneously
to Jordan form. Unlike the similarity transformations considered so far, however, this one 
converts the fundamental fusion matrix $Y$ into a {\em non-canonical} Jordan form.

Based on the Jordan chain (\ref{Yw}), we introduce the vectors
\be
 \wt_b^{(1)}\ =\ \mu_{b,1}^{(1)}w_b^{(1)},\qquad
 \wt_b^{(2)}\ =\ \mu_{b,2}^{(2)}w_b^{(2)}+ \mu_{b,1}^{(2)}w_b^{(1)},\qquad
 \wt_b^{(3)}\ =\ \mu_{b,3}^{(3)}w_b^{(3)}+ \mu_{b,2}^{(3)}w_b^{(2)}+ \mu_{b,1}^{(3)}w_b^{(1)}
\ee
where $\mu_{b,\ell}^{(\ell)}\neq0$. 
Unlike the original triplet $w_b^{(1)},w_b^{(2)},w_b^{(3)}$, the triplet 
$\wt_b^{(1)},\wt_b^{(2)},\wt_b^{(3)}$ does {\em not} form a Jordan chain. 
Instead, it forms the {\em generalized} Jordan chain
\be
 Y\big(\wt_b^{(1)}\ \wt_b^{(2)}\ \wt_b^{(3)}\big)\ =\ \big(\wt_b^{(1)}\ \wt_b^{(2)}\ \wt_b^{(3)}\big)
  \left(\!\!\begin{array}{ccc} \beta_b&\frac{\mu_{b,2}^{(2)}}{\mu_{b,1}^{(1)}}&
      \frac{\mu_{b,2}^{(2)}\mu_{b,2}^{(3)}
          -\mu_{b,1}^{(2)}\mu_{b,3}^{(3)}}{\mu_{b,1}^{(1)}\mu_{b,2}^{(2)}} 
  \\
   0&\beta_b&\frac{\mu_{b,3}^{(3)}}{\mu_{b,2}^{(2)}}
  \\ 
   0&0&\beta_b \end{array}\!\!\right)
\label{Yww3x3}
\ee
This generalization is particularly useful if the $(3\times3)$-matrix in (\ref{Yww3x3}) 
can be written as
\be
 \phi(\Jc_{\theta_b,3})\ =\ \phi(\!\left(\!\!\begin{array}{ccc} \theta_b&1&0
     \\ 0&\theta_b&1 \\ 0&0&\theta_b \end{array}\!\!\right)\!)
   \ =\ \left(\!\!\begin{array}{ccc} \phi(\theta_b)&\phi'(\theta_b)&\frac{1}{2}\phi''(\theta_b)
     \\ 0&\phi(\theta_b)&\phi'(\theta_b) \\ 0&0&\phi(\theta_b) \end{array}\!\!\right)
   \ =\ \left(\!\!\begin{array}{ccc} \beta_b&-2\sin\theta_b&-\cos\theta_b
     \\ 0&\beta_b&-2\sin\theta_b \\ 0&0&\beta_b \end{array}\!\!\right)
\label{phiJ}
\ee
where
\be
 \phi(\theta)\ =\ 2\cos\theta
\ee

We note that setting 
\be
 \mu_{b,\ell}^{(\ell)}\ =\ \frac{(-1)^{b-1}(-2\sin\theta_b)^{\ell}}{p\sqrt{2p}},\qquad
  \mu_{b,1}^{(2)}\ =\ \frac{2(-1)^b\cos\theta_b}{p\sqrt{2p}},\qquad
   \mu_{b,\kappa}^{(3)}\ =\ \frac{(2\kappa-1)(-1)^{b-1}\sin\kappa\theta_b}{p\sqrt{2p}}
\ee
respects (\ref{phiJ}).
In this case, and written in the basis (\ref{order}) indicated by the labeling $\ketw{\kappa,s},
\ketw{\R_\kappa^b}$, the entries of the vectors $\wt_b^{(\ell)}$ read
\bea
 \wt_{b'}^{(1)}\ =\ \left(\!\!\!\begin{array}{c}  \frac{2(-1)^{\kappa b'}\sin s\theta_{b'}}{p\sqrt{2p}}
   \\[.3cm] \hline\\[-.3cm]  0
  \end{array}\!\!\!\right),\quad
 \wt_{b'}^{(2)}\ =\ \left(\!\!\!\begin{array}{c}  \frac{2(-1)^{\kappa b'}s\cos s\theta_{b'}}{p\sqrt{2p}}
   \\[.3cm] \hline\\[-.3cm]  \frac{4(-1)^{(\kappa-1)b'}p\cos b\theta_{b'}}{p\sqrt{2p}}
  \end{array}\!\!\!\right),\quad
  \wt_{b'}^{(3)}\ =\ \left(\!\!\!\begin{array}{c}  \frac{(-1)^{\kappa b'+1}s^2\sin s\theta_{b'}}{p\sqrt{2p}}
   \\[.3cm] \hline\\[-.3cm]  \frac{4(-1)^{\kappa b'-b'+1}pb\sin b\theta_{b'}}{p\sqrt{2p}}
  \end{array}\!\!\!\right)
\eea
where there are $2p$ entries above and $2p-2$ entries below the horizontal separator
in a given vector.
It is likewise convenient to normalize the eigenvectors $v_{(\kappa'-1)p}$ and $v_b$
by introducing the vectors
\bea
 \vt_{(\kappa'-1)p}&=&\mu_{(\kappa'-1)p}v_{(\kappa'-1)p}
   \ =\ \frac{(-1)^{(\kappa'-1)(p-1)}}{p\sqrt{2p}}v_{(\kappa'-1)p}
   \ =\ \left(\!\!\!\begin{array}{c}  \frac{s(-1)^{(\kappa'-1)(\kappa p+s)}}{p\sqrt{2p}}
   \\[.3cm] \hline\\[-.3cm]  \frac{2p(-1)^{(\kappa'-1)((\kappa-1)p+b)}}{p\sqrt{2p}}
  \end{array}\!\!\!\right)\nn
 \vt_{b'}&=&\mu_{b'}v_{b'}\ =\ \frac{2(-1)^{b'}\sin\theta_{b'}}{p\sqrt{2p}}v_{b'}
     \ =\ \left(\!\!\!\begin{array}{c}  \frac{2(-1)^{\kappa b'+\kappa-1}\sin s\theta_{b'}}{p\sqrt{2p}}
   \\[.3cm] \hline\\[-.3cm]  0
  \end{array}\!\!\!\right)
\eea

Now, constructed by concatenating the vectors $\vt_j$ and $\wt_b^{(\ell)}$, the similarity matrix
\be
 \Qt\ =\ \left(\!\!\begin{array}{c|cccc|c|cccc|c} \vt_0&\vt_1&\wt_1^{(1)}&\wt_1^{(2)}&\wt_1^{(3)}
    &\ldots&
    \vt_{p-1}&\wt_{p-1}^{(1)}&\wt_{p-1}^{(2)}&\wt_{p-1}^{(3)}&\vt_p  \end{array}\!\!\right)
\label{Qtvw}
\ee
converts $X$ and $Y$ simultaneously 
\be
 \Qt^{-1}X\Qt\ =\ \Jt_X,\qquad\quad  \Qt^{-1}Y\Qt\ =\ \Jt_Y
\ee
into the Jordan forms 
\be
 \Jt_X\ =\ J_X,\qquad\quad
 \Jt_Y\ =\ \mathrm{diag}\big(\beta_0;\beta_1,\phi(\Jc_{\theta_1,3});\ldots;
     \beta_{p-1},\phi(\Jc_{\theta_{p-1},3});\beta_p\big)
\ee
It is noted that $\Jt_Y$ is a non-canoncal Jordan form, as already announced.
More generally, for $\Nc=X^{\kappa-1}f(Y)$ as in (\ref{NXf}), we have
\be
 \Qt^{-1}\Nc\Qt\ =\ \Jt_{\Nc}
    \ =\ J_X^{\kappa-1}f\circ\phi\big(\mathrm{diag}\big(\theta_0;\ldots;\theta_b,
    \Jc_{\theta_b,3};\ldots;\theta_p\big)\big)
\ee
with $b$ running from $1$ to $p-1$.
That is,
\be
 \Jt_{\Nc}\ =\ \mathrm{diag}\big(f(\beta_0);\ldots;(-1)^{(\kappa-1)(b-1)}f(\beta_b),
    (-1)^{(\kappa-1)b}f\circ\phi(\Jc_{\theta_b,3});\ldots;(-1)^{(\kappa-1)p}f(\beta_p)\big)
\label{JtN}
\ee
where
\bea
 f\circ\phi(\Jc_{\theta_b,3})&=&\left(\!\!\begin{array}{ccc} f\circ\phi(\theta_b)&(f\circ\phi)'(\theta_b)
   &\hf(f\circ\phi)''(\theta_b)
     \\ 0&f\circ\phi(\theta_b)&(f\circ\phi)'(\theta_b) \\ 0&0&f\circ\phi(\theta_b) \end{array}\!\!\right)\nn
 &=&\left(\!\!\begin{array}{ccc} f(\beta_b)&-2\sin\theta_bf'(\beta_b)
   &-\cos\theta_bf'(\beta_b)+2\sin^2\theta_bf''(\beta_b)
     \\ 0&f(\beta_b)&-2\sin\theta_bf'(\beta_b) \\ 0&0&f(\beta_b) \end{array}\!\!\right)
\label{fphiJ}
\eea

It is straightforward, albeit somewhat tedious, to show that the inverse of $\Qt$ is given by
\be
 \Qt^{-1}\ =\ \frac{1}{p\sqrt{2p}}\left(\!\!\!\begin{array}{c|c|c}
  0&p&p-b
 \\ \hline 
  \vdots & \vdots & \vdots
 \\ \hline 
  (-1)^{\kappa(b'-1)+1}p^2\sin b\theta_{b'}&0&\frac{1}{4}(-1)^{(\kappa-1)(b'-1)}p^2\sin b\theta_{b'}
 \\
  (-1)^{\kappa b'}p^2\sin b\theta_{b'}&0&\frac{1}{4}(-1)^{(\kappa-1)b'+1}\big(p^2-2(p-b)^2\big)
    \sin b\theta_{b'}
 \\
  0&(-1)^{(\kappa-1)b'}p&(-1)^{(\kappa-1)b'}(p-b)\cos b\theta_{b'}
 \\
  0&0&(-1)^{(\kappa-1)b'+1}\sin b\theta_{b'}
 \\ \hline
  \vdots & \vdots & \vdots
 \\ \hline 
  0&(-1)^{(\kappa-1)p}p&(-1)^{(\kappa-1)p+b}(p-b)
  \end{array}\!\!\!\right)
\label{Qtinv}
\ee
The columns are labeled by $\ketw{\kappa,b}$, $\ketw{\kappa,p}$ and $\ketw{\R_{\kappa}^{b}}$,
while the rows are labeled by $\vt_0$, the $p-1$ quadruplets $\vt_{b'}$, $\wt_{b'}^{(1)}$, 
$\wt_{b'}^{(2)}$ and $\wt_{b'}^{(3)}$, and $\vt_p$.
Finally, the determinant of $\Qt$ follows from that of $Q$ (\ref{detQ}) and is found to be
\be
 \mathrm{det}\Qt\ =\ \mu_0\mu_p\Big(\prod_{b=1}^{p-1}\mu_{b}\prod_{\ell=1}^3\mu_{b,\ell}^{(\ell)}
    \Big)\mathrm{det}Q\ =\ 
  \frac{\mathrm{det}Q}{2^{2p-1}p^{6p-10}}\ =\ (-1)^p\big(\frac{8}{p}\big)^{p-1}
\label{detQt}
\ee

\section{Generalized Verlinde formula}
\label{SecVerlinde}

\subsection{Characters and modular data}

The irreducible characters are given by
\be
  \chih_{\kappa,s}(q)\ =\ \chit[\ketw{\kappa,s}](q)\ =\ 
     \frac{1}{\eta(q)}\sum_{j\in\mathbb{Z}}(2j+\kappa)q^{p(j+\frac{\kappa p-s}{2p})^2}
\label{Wirrchar}
\ee
where $\eta(q)$ is the Dedekind eta function
\be
 \eta(q)\ =\ q^{1/24}\prod_{m=1}^\infty(1-q^m)
\label{eta}
\ee
The characters of the indecomposable rank-2 representations are given by
\be
 \chit[\ketw{\R_{1}^{b}}](q)\ =\ \chit[\ketw{\R_{2}^{p-b}}](q)\ =\ 
  2\chih_{2,b}(q)+2\chih_{1,p-b}(q)
\label{chiR}
\ee
There are $p+1$ linearly independent projective characters, namely 
$\chih_{\kappa,p}(q)$ for $\kappa\in\mathbb{Z}_{1,2}$ and the ones in (\ref{chiR}).

The set of irreducible characters (\ref{Wirrchar})
does not close under modular transformations. Instead, a representation of the 
modular group is obtained~\cite{Flo9509,Flo9605,FG0509,FGST0606b} by enlarging
the set with the $p-1$ so-called {\em pseudo-characters}
\be
 \chih_{0,b}(q)\ =\ i\tau\big(b\chih_{1,p-b}(q)-(p-b)\chih_{2,b}(q)\big)
\label{pseudo}
\ee
where the modular parameter is given by
\be
 q\ =\ e^{2\pi i\tau}
\label{qtau}
\ee
Writing the associated (generalized) modular $S$-matrix in block form with respect to 
the distinction between proper characters $\chih_{\kappa,s}(q)$ and pseudo-characters 
$\chih_{0,b}(q)$, the entries read
\be
 S\ =\ \begin{pmatrix}S_{\kappa,s}^{\kappa'\!,s'}&S_{\kappa,s}^{0,b'}\\[4pt]
  S_{0,b}^{\kappa'\!,s'}&S_{0,b}^{0,b'}\end{pmatrix}\ =\
  \begin{pmatrix}\frac{(2-\delta_{s',p})(-1)^{\kappa s'+ \kappa' s+ \kappa \kappa' p}
     s\cos\frac{ss'\pi}{p}}{p\sqrt{2p}}&
  \frac{2(-1)^{\kappa b'}\sin\frac{sb'\pi}{p}}{p\sqrt{2p}}\\[6pt]
  \frac{2(-1)^{\kappa' b}(p-s')\sin\frac{bs'\pi}{p}}{\sqrt{2p}}&0\end{pmatrix}
\label{Sks}
\ee
Here the lower (upper) indices refer to the row (column) labeling.
This matrix is not symmetric and not unitary, but satisfies $S^2=I$.
We note that
\be
 S_{\kappa,s}^{1,p-b}\ =\ S_{\kappa,s}^{2,b}
\ee
implying that, under the modular transformation $\tau\to\frac{-1}{\tau}$, the $2p$ irreducible
characters transform into linear combinations of the $p+1$ projective characters (with expansion 
coefficients $S_{\kappa,s}^{\kappa',p}$ and $\half S_{\kappa,s}^{2,b}$) and the $p-1$ 
pseudo-characters (with expansion coefficients $S_{\kappa,s}^{0,b}$), only.
We also introduce
\be
 S_{\ketw{\R_{\kappa}^b}}^{\kappa'\!,s'}
   \ =\ 2\big(S_{\kappa,p-b}^{\kappa'\!,s'}+S_{2\cdot\kappa,b}^{\kappa'\!,s'}\big)
\ee
and similarly for related combinations.
We finally note that, formally,
\be
 S_{\kappa,s}^{2,b}\ =\ \frac{\pa}{\pa\theta_b}S_{\kappa,s}^{0,b}
\label{deriv}
\ee

Alternatively, one can introduce the $2p$-dimensional, $\tau$-dependent 
(and thus {\em improper}) $S$-matrix 
\be
 \Sc-i\tau\Sct
 \label{Sc}
\ee
(here written in calligraphic to distinguish it from the proper $S$-matrix in (\ref{Sks}))
obtained by expanding the pseudo-characters in terms of the irreducible characters.
Its entries thus read
\be
 \Sc_{\kappa,s}^{\kappa'\!,s'}\ =\ S_{\kappa,s}^{\kappa'\!,s'},\qquad
 \Sct_{\kappa,s}^{\kappa'\!,s'}\ =\ \frac{2(-1)^{\kappa s'+\kappa's+\kappa\kappa' p}(p-s')
   \sin\frac{ss'\pi}{p}}{p\sqrt{2p}}
\label{S0S1}
\ee
from which it follows that
\be
 (p-b)\Sct_{\kappa,s}^{1,p-b}\ =\ -b\Sct_{\kappa,s}^{2,b} 
\ee
and
\be
 \Sct_{\kappa,s}^{1,b'}\ =\ -(p-b')S_{\kappa,s}^{0,p-b'},\qquad
 \Sct_{\kappa,s}^{2,b'}\ =\ (p-b')S_{\kappa,s}^{0,b'},\qquad
 \Sct_{\kappa,s}^{\kappa'\!,p}\ =\ 0
\ee
It is easily seen that
an expression can be written in terms of the proper $S$-matrix $S$ if and only if it can be 
written in terms of the improper $S$-matrix $\Sc-i\tau\Sct$. We use exclusively
the proper $S$-matrix $S$ in the following.

\subsection{Generalized Verlinde formula}
\label{SecGenVerFor}

The objective here is to express the spectral decompositions of the fusion
matrices in terms of the modular data. The various expressions are {\em not} unique,
as indicated by the rather trivial identity
$S_{2,1}^{\kappa,p}/S_{1,1}^{1,p}=S_{\kappa,1}^{0,1}/S_{1,1}^{0,1}$, for example. 
First, we observe that the eigenvalues can be written as  
\be
 \beta_{(\kappa'-1)p}\ =\ \frac{S_{1,2}^{\kappa',p}}{S_{1,1}^{\kappa',p}},\qquad\quad
 \beta_{b'}\ =\ \frac{S_{1,2}^{0,b'}}{S_{1,1}^{0,b'}}
\ee
and that
\be
  f_s(\beta_{(\kappa'-1)p})\ =\ \frac{S_{1,s}^{\kappa',p}}{S_{1,1}^{\kappa',p}},\quad
    f_s(\beta_{b'})\ =\ \frac{S_{1,s}^{0,b'}}{S_{1,1}^{0,b'}},\quad
  f_{p+b}(\beta_{(\kappa'-1)p})\ =\ 
    \frac{S_{\ketw{\R_1^b}}^{\kappa',p}}{S_{1,1}^{\kappa',p}},
        \quad  
    f_{p+b}(\beta_{b'})\ =\ \frac{S_{\ketw{\R_1^b}}^{0,b'}}{S_{1,1}^{0,b'}}
\ee
With reference to (\ref{fphiJ}), the remaining entries of the Jordan form 
$\Jt_{\Nc}$ (\ref{JtN}) follow from
\bea
 -2\sin\theta_{b'}f_s'(\beta_{b'})
   &=&\frac{S_{1,1}^{0,b'}S_{1,s}^{2,b'}-S_{1,1}^{2,b'}S_{1,s}^{0,b'}}{(S_{1,1}^{0,b'})^2}\nn
 -2\sin\theta_{b'}f_{p+b}'(\beta_{b'})
   &=&\frac{S_{1,1}^{0,b'}S_{\ketw{\R_1^b}}^{2,b'}
      -S_{1,1}^{2,b'}S_{\ketw{\R_1^b}}^{0,b'}}{(S_{1,1}^{0,b'})^2}
    \ =\ \frac{S_{\ketw{\R_1^b}}^{2,b'}}{S_{1,1}^{0,b'}}\nn
 -\cos\theta_{b'}f_s'(\beta_{b'})+2\sin^2\theta_{b'}f''_s(\beta_{b'})
  &=&\frac{S_{1,s}^{1,p}S_{1,s}^{1,p}S_{1,s}^{0,b'}}{S_{2,1}^{2,p}S_{1,2}^{1,p}S_{1,1}^{0,b'}}
    +\frac{S_{1,1}^{1,p}S_{1,s}^{0,b'}}{S_{1,2}^{1,p}S_{1,1}^{0,b'}}
    +\frac{S_{2,1}^{2,p}S_{1,2}^{0,b'}S_{2,s}^{2,b'}+(S_{1,1}^{2,b'})^2
       S_{1,s}^{0,b'}}{(S_{1,1}^{0,b'})^3}
  \nn
 -\cos\theta_{b'}f_{p+b}'(\beta_{b'})+2\sin^2\theta_{b'}f''_{p+b}(\beta_{b'})
  &=&\frac{S_{1,1}^{2,p}S_{1,2}^{1,p}S_{1,b'}^{2,p}S_{1,b}^{0,b'}}{(S_{1,1}^{1,p})^3S_{1,1}^{0,b'}}
   -\frac{S_{1,1}^{2,b'}S_{\ketw{\R_1^b}}^{2,b'}}{(S_{1,1}^{0,b'})^2}
\eea
The columns of the similarity matrix $\Qt$ (\ref{Qtvw}) can be written as
\bea
 &&\qquad\qquad\qquad
  \vt_{(\kappa'-1)p}\ =\ \left(\!\!\!\begin{array}{c}  S_{\ketw{\kappa,s}}^{\kappa',p}
   \\[.3cm] \hline\\[-.3cm]  S_{\ketw{\R_{\kappa}^{b}}}^{\kappa',p}
  \end{array}\!\!\!\right),\qquad
 \vt_{b'}\ =\ 
    \left(\!\!\!\begin{array}{c}  \frac{S_{2,1}^{\kappa,p}}{S_{1,1}^{1,p}}S_{\ketw{\kappa,s}}^{0,b'}
   \\[.3cm] \hline\\[-.3cm]  \frac{S_{2,1}^{\kappa,p}}{S_{1,1}^{1,p}}S_{\ketw{\R_{\kappa}^{b}}}^{0,b'}
  \end{array}\!\!\!\right)
 \nn
 &&\wt_{b'}^{(1)}\ =\ \left(\!\!\!\begin{array}{c}  S_{\ketw{\kappa,s}}^{0,b'}
   \\[.3cm] \hline\\[-.3cm]  S_{\ketw{\R_{\kappa}^{b}}}^{0,b'}
  \end{array}\!\!\!\right),\qquad
 \wt_{b'}^{(2)}\ =\ \left(\!\!\!\begin{array}{c}  S_{\ketw{\kappa,s}}^{2,b'}
   \\[.3cm] \hline\\[-.3cm]  S_{\ketw{\R_{\kappa}^{b}}}^{2,b'}
  \end{array}\!\!\!\right),\qquad
 \wt_{b'}^{(3)}\ =\ \left(\!\!\!\begin{array}{c}  
   \frac{S_{1,s}^{1,p}S_{1,s}^{1,p}}{S_{2,1}^{2,p}S_{1,2}^{1,p}}S_{\ketw{\kappa,s}}^{0,b'}
   \\[.3cm] \hline\\[-.3cm]  
    \frac{S_{1,1}^{2,p}S_{1,2}^{1,p}S_{1,b'}^{2,p}}{(S_{1,1}^{1,p})^3}
     S_{\ketw{\kappa,b}}^{0,b'}
    \end{array}\!\!\!\right) \qquad
\eea
while the entries of $\Qt^{-1}$ (\ref{Qtinv}) follow from
\bea
 &&
  \frac{p}{p\sqrt{2p}}\ =\ S^{1,p}_{\kappa,p},\qquad
    \frac{p-b}{p\sqrt{2p}}\ =\ S^{1,p}_{\kappa,p-b},\qquad
  \frac{(-1)^{\kappa(b'-1)+1}p^2\sin b\theta_{b'}}{p\sqrt{2p}}
  \ =\ \frac{S^{1,p}_{1,p}S^{1,p}_{1,p}}{S^{1,p}_{1,2}S^{\kappa,p}_{2,1}}S^{0,b'}_{\kappa,b}\nn
&& 
 \frac{\frac{1}{4}(-1)^{(\kappa-1)(b'-1)}p^2\sin b\theta_{b'}}{p\sqrt{2p}}
  \ =\ \frac{S^{\kappa,p}_{2,1}\big(S^{1,p}_{1,p}\big)^2}{\big(S^{1,p}_{1,2}\big)^3}
    S^{0,b'}_{\kappa-1,b},\qquad
 \frac{(-1)^{\kappa b'}p^2\sin b\theta_{b'}}{p\sqrt{2p}}
  \ =\ \frac{S^{1,p}_{1,p}S^{1,p}_{1,p}}{S^{1,p}_{1,2}S^{1,p}_{1,1}}S^{0,b'}_{\kappa,b}\nn
 &&\frac{\frac{1}{4}(-1)^{(\kappa-1)b'+1}\big(p^2-2(p-b)^2\big)\sin b\theta_{b'}}{p\sqrt{2p}}
  \ =\ \frac{S^{1,1}_{1,p}\big(S^{1,p}_{1,p}\big)^2}{\big(S^{1,p}_{1,2}\big)^3}
    S^{0,b'}_{\kappa,p-b}
    +\frac{\big(S^{1,p}_{1,p-b}\big)^2}{\big(S^{1,p}_{1,2}\big)^2}
    S^{0,b'}_{\kappa-1,b}
  \nn
 &&\frac{(-1)^{(\kappa-1)b'}p}{p\sqrt{2p}}
      \ =\ \frac{S^{1,p}_{1,1}}{S^{1,p}_{1,2}}S_{\kappa,p}^{2,b'},\qquad
   \frac{(-1)^{(\kappa-1)b'}(p-b)\cos b\theta_{b'}}{p\sqrt{2p}}
     \ =\ \frac{S_{1,1}^{1,p}}{S_{1,2}^{1,p}}S_{\kappa,p-b}^{2,b'}
    \nn
 &&
  \frac{(-1)^{(\kappa-1)b'+1}\sin b\theta_{b'}}{p\sqrt{2p}}
     \ =\ \frac{S_{1,1}^{1,p}}{S_{1,2}^{1,p}}S_{\kappa,p-b}^{0,b'},\quad
  \frac{(-1)^{(\kappa-1)p}p}{p\sqrt{2p}}\ =\ S^{2,p}_{\kappa,p},\quad
  \frac{(-1)^{(\kappa-1)p+b}(p-b)}{p\sqrt{2p}}\ =\ S^{2,p}_{\kappa,p-b}
  \nn
\eea
In summary, the announced generalized Verlinde formula reads
\be
 \Nc\ =\ \Qt\Jt_{\Nc}\Qt^{-1}
\ee
where $\Qt$, $\Jt_{\Nc}$ and $\Qt^{-1}$ are expressed in terms of the modular data
as outlined above.

\section{Partition functions}
\label{SecPartition}

Due to the presence of reducible yet indecomposable representations, the fusion algebra
contains more information than needed for the computation of partition functions as the
latter are given in terms of characters only. It is therefore natural to try to identify reductions
of the fusion algebra which can replace it when considering partition functions.
It is the objective here to outline how certain rings of equivalence classes of 
fusion-algebra generators do the job. As we will see, two simple requirements
ensure that the partition functions can be expressed in terms of the ring data.

In a given (possibly logarithmic) CFT, we let $\{\chit_i(q)\}$
denote the set of irreducible characters and $\{F_\mu\}$
the set of generators of the fusion algebra
\be
 F_\mu\otimes F_\nu\ =\ \bigoplus_\la {N_{\mu,\nu}}^\la F_\la,\qquad\quad
  {N_{\mu,\nu}}^\la\in\mathbb{N}_0
\label{FFNF}
\ee 
The character of a fusion generator is obtained by acting on it with $\chit$
\be
 \chit[F_\mu](q):=\ \sum_i {f_\mu}^i\chit_i(q),\qquad\quad {f_\mu}^i\in\mathbb{N}_0
\label{Ff}
\ee
This map extends by linearity. We refer to the matrix formed by the coefficients 
${f_\mu}^i$ as the {\em structure} matrix.

We let $\{G_m\}$
denote the set of equivalence classes of the linear span of fusion generators with respect to some
equivalence relation $\sim$. The projector onto these classes is denoted by $G$ and maps
a fusion generator into a linear combination of equivalence classes
\be
 G[F_\mu]:=\ \sum_m {h_\mu}^mG_m,\qquad\quad {h_\mu}^m\in\mathbb{C}
\label{GFhG}
\ee
This projector extends by linearity with respect to direct sums. 

Let a multiplication $\ast$ be defined on the set of equivalence classes 
\be
 G_m\ast G_n\ =\ \sum_\ell {M_{m,n}}^\ell G_\ell,\qquad\quad {M_{m,n}}^\ell\in\mathbb{C}
\ee
For this to be compatible with the fusion rules, we {\em require} that 
\be
 G[F_\mu\otimes F_\nu]\ =\ G[F_\mu]\ast G[F_\nu]
\label{GFFGFGF}
\ee
and subsequently say that the fusion rules (\ref{FFNF}) induce the multiplication rules on 
the equivalence classes. As a consequence, we have
\be
 \sum_\la {N_{\mu,\nu}}^\la {h_\la}^\ell\ =\ 
  \sum_{m,n}{h_\mu}^m {h_\nu}^n {M_{m,n}}^\ell
\ee

We also introduce a map $\tilde{\chit}$ from the set of equivalence classes to the set of characters
\be
 \tilde{\chit}[G_m]:=\ \sum_i {g_m}^i\chit_i(q),\qquad\quad {g_m}^i\in\mathbb{C}
\ee
{\em requiring} that
\be
 \tilde{\chit}\circ G\ =\ \chit
\ee
This implies that every lift $G^{-1}_m$ of the equivalence class $G_m$ to the 
set of fusion generators has the same character
\be
 \chit[G^{-1}_m]\ =\ \tilde{\chit}[G_m]
\ee
{}From examining $\chit[F_\mu]$, it follows that
\be
 {f_\mu}^i\ =\ \sum_m {h_\mu}^m {g_m}^i
\ee

We now consider the partition function (to be discussed further in Section~\ref{SecPart})
\be
 Z_{\mu,\nu}(q):=\ \chit[F_\mu\otimes F_\nu](q)
  \ =\ \sum_{\la,i}{N_{\mu,\nu}}^\la {f_\la}^i \chit_i(q)
\label{Zab}
\ee
Using the above, including the two requirements, we see that this can be written in terms of 
the data for the equivalence classes
\be
 Z_{\mu,\nu}(q)\ =\ \sum_{m,n,\ell,i}{h_\mu}^m {h_\nu}^n {M_{m,n}}^\ell {g_\ell}^i \chit_i(q)
\label{Zab2}
\ee
By construction, we thus have
\be 
 \sum_{m,n,\ell}{h_\mu}^m {h_\nu}^n {M_{m,n}}^\ell {g_\ell}^i\in\mathbb{N}_0
\label{ringint}
\ee
It follows that, in order to obtain the partition functions (\ref{Zab}), 
it suffices to know the algebra of the equivalence classes provided this algebra
respects the two requirements. As already mentioned, this property of the partition functions is
the rationale for imposing the two requirements.
As we will discuss in the following, when the equivalence classes correspond to the 
generators of the Grothendieck group of the characters of ${\cal WLM}(1,p)$,
the induced multiplication rules of the corresponding Grothendieck ring
follow straightforwardly from the fusion algebra, and $\tilde\chit$ is an almost trivial bijection.

A natural objective is to determine a minimal algebra, that is, one of smallest possible 
dimension, of equivalence classes compatible with the fusion algebra. 
A lower bound for this dimension is the number of linearly independent
characters appearing in the fusion algebra, where we note that this number
can be smaller that the number of irreducible characters.
Another interesting problem is to determine the minimal algebra
corresponding to a ring over the integers $\mathbb{Z}$. 
The fusion algebra itself is such a ring with $\ast=\!\fus\!$, so an upper bound on the
dimension is given, in this case, by the dimension of the fusion algebra.
We hope to address these questions elsewhere.

\subsection{Grothendieck ring of ${\cal WLM}(1,p)$}

The Grothendieck ring of ${\cal WLM}(1,p)$ is obtained by elevating the
character identities of ${\cal WLM}(1,p)$ to identities between the corresponding fusion
generators. From (\ref{chiR}), we thus impose the equivalence relations
\be
  \ketw{\R_{1}^{b}}\ \sim\ \ketw{\R_{2}^{p-b}}\ \sim\ 
  2\ketw{2,b}\oplus2\ketw{1,p-b},\qquad\quad b\in\mathbb{Z}_{1,p-1}
\label{Gequiv}
\ee
In terms of equivalence classes, this means that $|\{G_m\}|=2p$ where
\be
 G[\ketw{\kappa,s}]\ =\ G_{\kappa,s},\qquad \quad
 G[\ketw{\R_1^b}]\ =\ G[\ketw{\R_2^{p-b}}]\ =\ 2G_{2,b}+2G_{1,p-b}
\label{GGGGGG}
\ee
It is easily verified that (\ref{GFFGFGF}) is respected by the multiplication $\ast$ 
whose multiplication rules are given by
\be
  G_{\kappa,s}\ast G_{\kappa',s'}\ =\ 
  \!\sum_{j=|s-s'|+1,\ \!\mathrm{by}\ \!2}^{p-|p-s-s'|-1}
  \!\!\! G_{\kappa\cdot \kappa',j}
    +\!\!\sum_{\beta=\eps(s+s'-p-1),\ \!\mathrm{by}\ \!2}^{s+s'-p-1}
\!\!\!(2-\delta_{\beta,0})\big(G_{\kappa\cdot\kappa',p-\beta}+G_{2\cdot\kappa\cdot\kappa',\beta}\big)   
\label{Gmult}
\ee
where $G_{\kappa,0}\equiv0$.
These rules actually correspond to a transcription of the first fusion rule in (\ref{fus}).
As already indicated, the map $\tilde\chit$ is simply given by the bijection
\be
 \tilde\chit[G_{\kappa,s}]\ =\ \chit[\ketw{\kappa,s}]\ =\ \chih_{\kappa,s}
\label{chiGchih}
\ee
between the set of Grothendieck generators and the set of irreducible characters.
The two index sets $\{m\}$ and $\{i\}$ can therefore be identified, and we have
\be
 {g_m}^i\ =\ {\delta_m}^i\quad \Rightarrow\quad {f_\mu}^i\ =\ {h_\mu}^i
\ee

\subsection{Partition functions in ${\cal WLM}(1,p)$}
\label{SecPart}

From the lattice description~\cite{PRR0803} of ${\cal WLM}(1,p)$, 
every indecomposable representation 
appearing in (\ref{Jp}) can be associated with a boundary condition. 
The corresponding characters are in general not linearly independent.
We can nevertheless talk about partition functions arising when combining two such
boundary conditions as in (\ref{Zab}). 
Following the discussion above, we thus have
\be
 Z_{\mu,\nu}(q)
  \ =\ \sum_{\la,i}{N_{\mu,\nu}}^\la {f_\la}^i \chit_i(q)
  \ =\ \sum_{m,n,i}{h_\mu}^m {h_\nu}^n {M_{m,n}}^i \chit_i(q)
\label{ZZ}
\ee
implying that
\be
 \sum_{\la}{N_{\mu,\nu}}^\la {f_\la}^i
  \ =\ \sum_{m,n}{h_\mu}^m {h_\nu}^n {M_{m,n}}^i
\label{NM}
\ee
where the structure matrix
${f_\mu}^m={h_\mu}^m$ is the $\big((4p-2)\times 2p\big)$-dimensional matrix
defined by (\ref{Ff}) or equivalently by (\ref{GFhG}). The explicit form of this matrix follows
from (\ref{chiGchih}) and (\ref{chiR}) or equivalently from (\ref{GGGGGG}).
Ordering the rows as in (\ref{order}) and the columns as 
\be
 G_{1,1},G_{2,1};\ldots;G_{1,s},G_{2,s};\ldots;G_{1,p},G_{2,p}
\ee
the structure matrix is given by
\be
 {f_\mu}^m\ =\ {h_\mu}^m\ =\ \left(\!\!\begin{array}{ccc} I_{p-1}&0&0 \\ 0&I_{p-1}&0 \\ 0&0&I_2\\
    2C_{p-1}&2I_{p-1}&0 \\  2I_{p-1}&2C_{p-1} &0 
         \end{array}\!\!\right)
\label{fh}
\ee

\subsection{Relation between Verlinde formulas for ${\cal WLM}(1,p)$}

The generalized Verlinde formula derived in Section~\ref{SecGenVerFor} yields
the multiplicities ${N_{\mu,\nu}}^\la$ in (\ref{ZZ}), whereas the multiplicities
${M_{m,n}}^i$, also in (\ref{ZZ}), are given by the generalized Verlinde formulas for the
Grothendieck ring appearing in~\cite{GR0707,PRR0907}.
Here we demonstrate how the so-called Moore-Penrose inverse of ${h_\mu}^m$
allows us to isolate ${M_{m,n}}^i$ from the relation (\ref{NM}).

First, we recall that for every $n\times m$ matrix $A$, there is
a unique matrix $A^\dagger$ satisfying the four Penrose equations (see~\cite{Pen55,BG03}, 
for example)
\be
 AA^\dagger A\ =\ A,\qquad A^\dagger AA^\dagger\ =\ A^\dagger,\qquad
  (AA^\dagger)^\ast\ = \ AA^\dagger,\qquad (A^\dagger A)^\ast\ =\ A^\dagger A
\ee
where $A^\ast$ denotes the conjugate transpose of $A$.
The matrix $A^\dagger$ is called the Moore-Penrose inverse, or pseudoinverse for short, 
of $A$. Clearly, $A^\dagger$ is an $m\times n$ matrix, and if $A$ is nonsingular, 
then $A^\dagger=A^{-1}$.
It also follows readily that $AA^\dagger$ and $A^\dagger A$ are projection matrices.
Furthermore, if $A$ has full column rank, then $A^\ast A$ is invertible and 
$A^\dagger=(A^\ast A)^{-1}A^\ast$ implying, in particular, that $A^\dagger A=I$.

Now, the structure matrix (\ref{fh}) has full {\em column} rank so 
\be
 \sum_{\mu}h_m^{\dagger\ \!\mu}{h_\mu}^n\ =\ {\delta_m}^n
\ee
It follows that
\be
 {M_{m,n}}^\ell
   \ =\ \sum_{\mu,\nu,\la}h_m^{\dagger\ \!\mu}h_n^{\dagger\ \!\nu}{N_{\mu,\nu}}^\la {h_\la}^\ell
   \ =\ \sum_{\mu,\nu,\la}f_m^{\dagger\ \!\mu}f_n^{\dagger\ \!\nu}{N_{\mu,\nu}}^\la {f_\la}^\ell
\ee
which expresses the output ${M_{m,n}}^\ell$ of the generalized Verlinde
formula for the Grothendieck ring in terms of fusion data.
Isolating ${N_{\mu,\nu}}^\la$, on the other hand, from (\ref{NM}) is not achieved
by application of the Moore-Penrose inverse of the structure matrix
simply because this structure matrix does not have full {\em row} rank so
\be
 \sum_{i}{f_\la}^i f_i^{\dagger\ \!\mu}\ \neq\ {\delta_\la}^\mu
\ee

\section{Conclusion}

We have described the graph fusion algebra of ${\cal WLM}(1,p)$. The corresponding 
fusion matrices (adjacency matrices) are mutually commuting, but in general not diagonalizable. 
Nevertheless, they can be simultaneously brought to Jordan form, albeit typically non-canonical
Jordan form, by a similarity transformation. The two fundamental fusion matrices are
simultaneously brought to Jordan canonical form by the similarity matrix $Q$.
For every fusion matrix $\Nc$, we have provided a modified $Q$-matrix $Q_\Nc$
converting $\Nc$ to Jordan canonical form. These Jordan canonical forms are given
explicitly and consist of Jordan blocks of rank 1, 2 or 3.
The various similarity transformations and Jordan forms can be expressed in terms of
modular data. This gives rise to a generalized Verlinde formula for the fusion matrices.
Its relation to the partition functions in the model is discussed in a general framework.
By application of a particular structure matrix and its Moore-Penrose inverse,
this Verlinde formula reduces to the Verlinde-like formula~\cite{PRR0907} for 
the associated Grothendieck ring.

We recall that fusion graphs have been instrumental in the classification of rational 
conformal field theories on the cylinder and on the torus.
It is our hope that the present work will be a step towards
extending these fundamental insights to the logarithmic conformal field theories.
First, though, one should extend our results to the general series of ${\cal W}$-extended
logarithmic minimal models ${\cal WLM}(p,p')$. In this direction, we have recently 
worked out the corresponding graph fusion algebras and determined their
spectral decompositions \cite{Ras0911}. The next objective is to determine
the associated Verlinde-like formulas which we hope to address elsewhere.

{}From Section~\ref{SecPartition}, we recall that the partition functions can be expressed
in terms of a ring of equivalence classes of fusion-algebra generators provided two
simple requirements are respected. As indicated following (\ref{ringint}), in particular,
there are many interesting problems related to these fusion-algebra compatible rings. 
The determination of a minimal such ring
or the classification of similar rings over the integers are natural examples.

\subsection*{Acknowledgments}
\vskip.1cm
\noindent
This work is supported by the Australian Research Council. 
The author thanks Paul A. Pearce and Philippe Ruelle for helpful discussions.

\appendix

\section{On the polynomial fusion ring in Section \ref{SecPolRing}}

\subsection{Simplification of isomorphism}
\label{AppIso}

Here we prove the simplification (\ref{iso2}) of the isomorphism (\ref{iso}).
Recalling the notation 
\be
 P_n(x)\ =\ (x^2-4)U_{n-1}^3(\tfrac{x}{2}),\qquad 
 P_{n,n'}(x,y)\ =\ \big(T_n(\tfrac{x}{2})-T_{n'}(\tfrac{y}{2})\big)
  U_{n-1}(\tfrac{x}{2})U_{n'-1}(\tfrac{y}{2})
\ee
from~\cite{Ras0812}, we consider
\bea
 U_{n-1}(\tfrac{x}{2})U_{n'-1}(\tfrac{y}{2})
   &\equiv&\big(1+((\tfrac{x}{2})^2-1)U_{n-1}^2(\tfrac{x}{2})\big)
     U_{n-1}(\tfrac{x}{2})U_{n'-1}(\tfrac{y}{2})\nn
  &=&T_n^2(\tfrac{x}{2})U_{n-1}(\tfrac{x}{2})U_{n'-1}(\tfrac{y}{2})\nn
  &\equiv&U_{n-1}(\tfrac{x}{2})T_{n'}^2(\tfrac{y}{2})U_{n'-1}(\tfrac{y}{2})\nn
  &=&U_{n-1}(\tfrac{x}{2})
   \big(1+((\tfrac{y}{2})^2-1)U_{n'-1}^2(\tfrac{y}{2})\big)
     U_{n'-1}(\tfrac{y}{2})
\eea
The first equivalence is modulo $P_n(x)$, the second equivalence is modulo $P_{n,n'}(x,y)$
(applied twice), while the two equalities follow from the identity
\be
 T_n^2(z)\ =\ 1+(z^2-1)U_{n-1}^2(z)
\ee
As an immediate consequence, we see that
\be
 0\ \equiv\ U_{n-1}(\tfrac{x}{2})\big((\tfrac{y}{2})^2-1\big)U_{n'-1}^3
   (\tfrac{y}{2})\qquad(\mathrm{mod}\ P_n(x),P_{n,n'}(x,y))
\ee
For $n=1$, in which case $U_{n-1}\big(\frac{x}{2}\big)$ is a non-vanishing constant, 
this implies that
\be
 P_{n'}(y)\ \equiv\ 0\qquad(\mathrm{mod}\ P_1(x),P_{1,n'}(x,y))
\ee

\subsection{Quotient polynomial ring conditions}
\label{AppQuotient}

Here we complete the verification of the quotient polynomial ring conditions in (\ref{iso}).
Since
\be
 \big(\Jc_b-\beta_bI\big),\big(\Jc_b-\beta_bI\big)^2\ \neq\ 0,\qquad
   \big(\Jc_b-\beta_bI\big)^3\ =\ 0,\qquad\qquad b\in\mathbb{Z}_{1,p-1}
\ee
where $\Jc_b=\Jc_{\beta_b,3}$ as in (\ref{JY}),
it follows from the Jordan canonical form $J_Y$ of $Y$ that the minimal polynomial of $Y$ is 
indeed given by $P_p(Y)$ in (\ref{charY}).
It also follows that the characteristic polynomial of $Y$ is given as in (\ref{charY}).
Due to (\ref{charX}) and the commutativity of $X$ and $Y$, 
the explicit verification of (\ref{iso}) is thus completed once we have established 
that $\tilde{P}_{1,p}(X,Y)=0$ which is equivalent to
\be
 J_XU_{p-1}\big(\frac{J_Y}{2}\big)\ =\ T_p\big(\frac{J_Y}{2}\big)U_{p-1}\big(\frac{J_Y}{2}\big)
\label{JXU}
\ee
Using (\ref{JX}) and (\ref{UT0}), the left-hand side reads
\bea
 J_XU_{p-1}\big(\frac{J_Y}{2}\big)&=&\mathrm{diag}\Big(U_{p-1}(\al_0);
    U_{p-1}(\al_1),-U_{p-1}\big(\frac{\Jc_1}{2}\big);\ldots;
    (-1)^{b-1}U_{p-1}(\al_b),(-1)^bU_{p-1}\big(\frac{\Jc_b}{2}\big);\ldots;\nn
  &&\qquad\qquad \qquad
      (-1)^{p-2}U_{p-1}(\al_{p-1}),(-1)^{p-1}U_{p-1}\big(\frac{\Jc_{p-1}}{2}\big);
        (-1)^pU_{p-1}(\al_p)\Big)\nn
  &=&\mathrm{diag}\Big(p;0,(-1)^1U_{p-1}\big(\frac{\Jc_1}{2}\big);\ldots;0,(-1)^{p-1}U_{p-1}
     \big(\frac{\Jc_{p-1}}{2}\big);-p\Big)
\label{JXULeft}
\eea
Likewise, the right-hand side of (\ref{JXU}) reads
\be
 T_p\big(\frac{J_Y}{2}\big)U_{p-1}\big(\frac{J_Y}{2}\big)
 \ =\ \mathrm{diag}\Big(p;0,T_p\big(\frac{\Jc_1}{2}\big)U_{p-1}\big(\frac{\Jc_1}{2}\big);\ldots;
    0,T_p\big(\frac{\Jc_{p-1}}{2}\big)U_{p-1}\big(\frac{\Jc_{p-1}}{2}\big);-p\Big)
\ee
Using
\be
 f\big(\frac{\Jc_b}{2}\big)\ =\ \left(\!\!\begin{array}{ccc} f(\al_b)&\hf f'(\al_b)&\frac{1}{8} f''(\al_b) \\[3pt]
   0&f(\al_b)&\hf f'(\al_b) \\[3pt]     0&0&f(\al_b)   
    \end{array}\!\!\!\right)
\label{fJb}
\ee
for polynomial $f$, we find
\be
 T_p\big(\frac{\Jc_b}{2}\big)U_{p-1}\big(\frac{\Jc_b}{2}\big)\ =\ 
   \left(\!\!\begin{array}{ccc} (-1)^b&0&\frac{1}{8}T_p''(\al_b) \\[3pt] 0&(-1)^b&0 \\[3pt] 0&0&(-1)^b 
     \end{array}\!\!\!\right)
   \left(\!\!\begin{array}{ccc} 0&\hf U_{p-1}'(\al_b)&\frac{1}{8}U_{p-1}''(\al_b) \\[3pt] 
      0&0& \hf U_{p-1}'(\al_b) \\[3pt] 0&0&0
     \end{array}\!\!\!\right)
 \ =\ (-1)^b U_{p-1}\big(\frac{\Jc_b}{2}\big)
\ee
and thus recover (\ref{JXULeft}). This completes the explicit
verification of $\tilde{P}_{1,p}(X,Y)=0$ and hence of the isomorphism (\ref{iso}) 
already established, using structural and algebraic arguments, in~\cite{Ras0812}.

\section{Properties of the functions $f_k(x)$}
\label{AppPropfk}

Here we derive and list some useful properties of the functions $f_k(x)$ defined in (\ref{f}).

\subsection{Recursion relations and special values}
\label{Appfk}

For $p>2$, the functions satisfy recursive relations allowing us to express 
$x f_k(x)$, for $k\in\mathbb{Z}_{1,2p-2}$, as
\bea
 f_2(x)&=&x f_1(x)  \nn
 f_{k-1}(x)+f_{k+1}(x)&=&x f_k(x),\qquad\quad k\in\mathbb{Z}_{2,p-1}\nn
 f_{p+1}(x)&=&x f_p(x)\nn
 2f_p(x)+f_{p+2}(x)&=&x f_{p+1}(x)\nn
 f_{k-1}(x)+f_{k+1}(x)&=&x f_k(x),\qquad\quad k\in\mathbb{Z}_{p+2,2p-2}
\label{ff}
\eea
It follows that
\be
 \begin{array}{rll}
 f_2'(x)&=\quad\!\!x f_1'(x)+f_1(x),\qquad\qquad\qquad\qquad\quad\ \ 
      f_2''(x)&=\quad\!\! xf_1''(x)+2f_1'(x)  \\[3pt]
 f_{k-1}'(x)+f_{k+1}'(x)&=\quad\!\! x f_k'(x)+f_k(x),\qquad\qquad\ \!
     f_{k-1}''(x)+f_{k+1}''(x)&=\quad\!\!xf_k''(x)+2f_k'(x)\\[3pt]
 f_{p+1}'(x)&=\quad\!\!x f_p'(x)+f_p(x),\qquad\qquad\qquad\qquad\ \
      f_{p+1}''(x)&=\quad\!\!x f_p''(x)+2f_p'(x)\\[3pt]
 2f_p'(x)+f_{p+2}'(x)&=\quad\!\!x f_{p+1}'(x)+f_{p+1}(x),\qquad\ \ \!
    2f_p''(x)+f_{p+2}''(x)&=\quad\!\!x f_{p+1}''(x)+2f_{p+1}'(x)\\[3pt]
 f_{k-1}'(x)+f_{k+1}'(x)&=\quad\!\!x f_k'(x)+f_k(x),\qquad\qquad
     f_{k-1}''(x)+f_{k+1}''(x)&=\quad\!\!x f_k''(x)+2f_k'(x)
 \end{array}
\label{fdiff}
\ee
with the conditions on $k$ adopted from (\ref{ff}).
It is noted that we have not included any relations involving $xf_{2p-1}(x)$ for general $x$. 
Instead, we focus on evaluations at $x=\beta_j$ for $j\in\mathbb{Z}_{0,p}$, where
it is recalled that $\beta_j=2\al_j$ with $\al_j$ defined in (\ref{Ua}).
We thus find that
\be
 \begin{array}{rll}
 2(-1)^if_p(\beta_j)+f_{2p-2}(\beta_j)&=\quad\!\!\beta_j f_{2p-1}(\beta_j),
    \qquad\qquad\qquad \qquad\ \ \!
   &j\in\mathbb{Z}_{1,p-1}  \ \mathrm{or}\ i=j\in\{0,p\}\\[3pt]
 2(-1)^bf_p'(\beta_b)+f_{2p-2}'(\beta_b)&=\quad\!\!
   \beta_b f_{2p-1}'(\beta_b)+f_{2p-1}(\beta_b),\qquad\quad\ 
    &b\in\mathbb{Z}_{1,p-1}\\[3pt]
 2(-1)^bf_p''(\beta_b)+f_{2p-2}''(\beta_b)&=\quad\!\!
   \beta_b f_{2p-1}''(\beta_b)+2f_{2p-1}'(\beta_b),\qquad\quad 
    &b\in\mathbb{Z}_{1,p-1}
 \end{array}
\label{f2p}
\ee
In establishing these relations, we use that
\be
 xf_{2p-1}(x)-f_{2p-2}(x)\ =\ 2T_p(\tfrac{x}{2})U_{p-1}(\tfrac{x}{2})
\ee
and that
\be
 U_{p-1}(\al_b)\ =\ \tfrac{1}{p}T_p'(\al_b)\ =\ 0,\qquad\quad 
   T_p(\al_j)\ =\ \cos\big(p\theta_j\big)\ =\ (-1)^j
\label{UT0}
\ee
recalling from (\ref{Ua}) that $\theta_j=j\pi/p$.

At special values, the evaluation of the functions $f_k$ and their derivatives can be simplified.
Some of these results are collected here for simple reference. Additional expressions
are found in Appendix~\ref{AppTrig}. We have
\be
 f_s(\pm2)\ =\ s(\pm1)^{s-1},\qquad f_{p+b'}(\pm2)\ =\ 2p(\pm1)^{p+b'-1},\qquad 
    f_p(\beta_b)\ =\ f_{p+b'}(\beta_b)\ =\ 0
\ee 
and
\be
 \begin{array}{llll} f_s(0)\ =\ 0,\quad &f_s'(0)\ =\ j(-1)^{j-1},\quad &f_s''(0)\ =\ 0,\qquad &s\ =\ 2j \\[3pt]
    f_s(0)\ =\ (-1)^j,\quad &f_s'(0)\ =\ 0,\quad &f_s''(0)\ =\ j(j+1)(-1)^{j-1} ,\qquad &s\ =\ 2j+1
 \end{array}
\ee
and for $p$ odd
\be
 \begin{array}{lll} f_{p+b'}(0)\ =\ 0,\quad &f_{p+b'}'(0)\ =\ b'(-1)^{\frac{p+1}{2}+i},
    \ \quad f_{p+b'}''(0)\ =\ 0,
     \quad &b'\ =\ 2i-1 \\[3pt]
    f_{p+b'}(0)\ =\ 2(-1)^{\frac{p-1}{2}+i},\quad &f_{p+b'}'(0)\ =\ 0,
      \ \quad f_{p+b'}''(0)\ =\ \frac{p^2+4i^2-1}{2}(-1)^{\frac{p+1}{2}+i},
     \qquad &b'\ =\ 2i
 \end{array}
\ee
and for $p$ even
\be
 \begin{array}{llll} f_{p+b'}(0)\ =\ 0,\quad &f_{p+b'}'(0)\ =\ 0,\quad &f_{p+b'}''(0)
   \ =\ b'p(-1)^{\frac{p}{2}+i},  \qquad &b'\ =\ 2i-1 \\[3pt]
    f_{p+b'}(0)\ =\ 0,\quad &f_{p+b'}'(0)\ =\ p(-1)^{\frac{p}{2}+i-1},\quad &f_{p+b'}''(0)\ =\ 0,
     \qquad &b'\ =\ 2i
 \end{array}
\ee

\subsection{Trigonometric expressions}
\label{AppTrig}

The Chebyshev polynomials and their derivatives
appearing in $Q$ are evaluated at trigonometric values
\be
 U_{n-1}(\cos\theta)\ =\ \frac{\sin n\theta}{\sin\theta},\qquad\quad U_{n-1}(\pm1)\ =\ n(\pm1)^{n-1}
\ee
We thus have
\bea
 f_s(\beta_{b'})&=&\frac{\sin s\theta_{b'}}{\sin\theta_{b'}},\qquad\quad
   f_{s}'(\beta_{b'})\ =\ \frac{\sin s\theta_{b'}\cos\theta_{b'}
    -s\cos s\theta_{b'}\sin\theta_{b'}}{2\sin^3\theta_{b'}}
  \nn
 f_s''(\beta_{b'})&=&\frac{\sin s\theta_{b'}\big(1+2\cos^2\theta_{b'}\big)-\frac{3}{2}s\cos s\theta_{b'}
   \sin2\theta_{b'}-s^2\sin s\theta_{b'}\sin^2\theta_{b'}}{4\sin^5\theta_{b'}}
\eea
and
\bea
 f_{p+b}(\beta_{b'})&=&0,\qquad\quad
 f_{p+b}'(\beta_{b'})\ =\ \frac{(-1)^{b'-1}p\cos b\theta_{b'}}{\sin^2\theta_{b'}}\nn
 f_{p+b}''(\beta_{b'})&=&\frac{(-1)^{b'-1}p\big(b\sin b\theta_{b'}\sin\theta_{b'}
   +\frac{3}{2}\cos b\theta_{b'}\cos\theta_{b'}\big)}{\sin^4\theta_{b'}}
\eea

\subsection{Classification of zeros}
\label{AppZeros}

For $k\in\mathbb{Z}_{1,2}$, we have
\be
 f_1'(x)\ =\ f_1''(x)\ =\ 0,\qquad\quad f_2'(x)\ =\ 1,\quad f_2''(x)\ =\ 0
\label{12}
\ee
while for $k=s\in\mathbb{Z}_{3,p}$, we have
\be
 f_s'(\beta_b)\ =\ 0\qquad\Leftrightarrow\qquad \al_b\ =\ 0,\quad s\ \mathrm{odd}\qquad
  \Leftrightarrow\qquad b\ =\ \frac{p}{2},\quad s\ =\ 2j+1,\quad j\in\mathbb{Z}_{1,\frac{p}{2}-1}
\label{fs}
\ee
and
\be
 f_s''(\beta_b)\ =\ 0\qquad \Leftrightarrow\qquad \al_b\ =\ 0,\quad s\ \mathrm{even}\qquad
  \Leftrightarrow\qquad b\ =\ \frac{p}{2},\quad s\ =\ 2j,\quad j\in\mathbb{Z}_{2,\frac{p}{2}}
\label{fss}
\ee
For $k=p+b'\in\mathbb{Z}_{p+1,2p-1}$, we have
\be
 f_{p+b'}'(\beta_b)\ =\ T_{b'}'(\al_b)U_{p-1}(\al_b)+T_{b'}(\al_b)U_{p-1}'(\al_b)\ =\ 
  \cos\frac{bb'\pi}{p}U_{p-1}'(\al_b)
\ee
{}From (\ref{fs}), it then follows that
\be
 f_{p+b'}'(\beta_b)\ =\ 0\qquad\Leftrightarrow\qquad \cos\frac{bb'\pi}{p}\ =\ 0
   \qquad\Leftrightarrow\qquad     2bb'\ =\ mp,\quad m\ \mathrm{odd}
\ee
Since $m$ is odd, the last identity implies that $2(b,p)$, where $(b,p)$ 
denotes the greatest common divisor of $b$ and $p$, is a divisor of $p$ and hence that
$\frac{b}{(b,p)}$ is odd. 
{}From
\be
 b'\frac{b}{(b,p)}2(b,p)\ =\ m\frac{p}{2(b,p)}2(b,p),\qquad\quad
  \Big(\frac{b}{(b,p)},\frac{p}{2(b,p)}\Big)\ =\ 1
\ee
it then follows that $b'$ is an odd multiple of $\frac{p}{2(b,p)}$.
Since $b'\in\mathbb{Z}_{1,p-1}$, we thus have
\be
 b'\ =\ \frac{(2j-1)p}{2(b,p)},\qquad j\in\mathbb{Z}_{1,(b,p)},\quad 
   \frac{p}{2(b,p)}\in\mathbb{Z}_{1,\frac{p}{2}}
\label{bb}
\ee
That is, for given $b$, $f_{p+b'}'(\beta_b)=0$ if and only if $b'$ is of the form (\ref{bb}) and
$\frac{p}{2(b,p)}\in\mathbb{Z}_{1,\frac{p}{2}}$. 
Due to the symmetric conditions $b,b'\in\mathbb{Z}_{1,p-1}$ and $\cos\frac{bb'\pi}{p}=0$,
we likewise have that, for given $b'$, 
\be
 f_{p+b'}'(\beta_b)\ =\ 0\qquad \Leftrightarrow\qquad
    b\ =\ \frac{(2j-1)p}{2(b',p)},\qquad j\in\mathbb{Z}_{1,(b',p)},\quad
     \frac{p}{2(b',p)}\in\mathbb{Z}_{1,\frac{p}{2}}
\label{b}
\ee
Finally, still for $k=p+b'\in\mathbb{Z}_{p+1,2p-1}$, we have
\be
 f_{p+b'}''(\beta_b)\ =\ T_{b'}'(\al_b)U_{p-1}'(\al_b)+\hf T_{b'}(\al_b)U_{p-1}''(\al_b)
\ee
and
\be
 f_{p+b'}''(\beta_b)\ =\ 0\qquad\Leftrightarrow\qquad \al_b\ =\ 0,\quad p+b'\ \mathrm{even}\qquad
   \Leftrightarrow\qquad b\ =\ \frac{p}{2},\quad b'\ =\ 2j,\quad j\in\mathbb{Z}_{1,\frac{p}{2}-1}
\ee
We observe that $f_k'(\beta_b)=f_k''(\beta)=0$ if and only if $k=1$.


\end{document}